\documentclass[preprint,12pt]{aastex}
\usepackage{natbib,epsfig}
\usepackage{graphicx}
\usepackage{multirow}
\usepackage{amsmath}

\newcommand{\g}{\gamma}
\newcommand{\3}{\sqrt{3}}
\newcommand{\bo}{breakout }
\newcommand{\ou}{outermost }
\newcommand{\lu}{luminosity }
\newcommand{\R}{R_{5}}

\newcommand{\M}{M_{5}}
\newcommand{\E}{E_{53}}
\newcommand{\Mej}{M_{ej,5}}

\newcommand{\gfh}{\widehat{\g}_f}
\newcommand{\gih}{\widehat{\g}_i}

\newcommand{\mh}{\widehat{m}}
\newcommand{\vh}{\widehat{v}}
\newcommand{\tauh}{\widehat{\tau}}
\newcommand{\dhat}{\widehat{d}}

\newcommand{\Eh}{\widehat{E}}
\newcommand{\Th}{\widehat{T}}
\newcommand{\Lh}{\widehat{L}}

\newcommand{\rh}{\widehat{r}}

\newcommand{\G}{\Gamma}
\newcommand{\Gu}{\Gamma_u}
\newcommand{\be}{\beta}
\newcommand{\x}{x_{\pm,\gamma}}
\newcommand{\ngus}{n_{\gamma \rightarrow{us}}}

\begin{document}

\title{Relativistic shock breakouts - a variety of gamma-ray flares: from low luminosity gamma-ray bursts to type Ia supernovae}
\author{Ehud Nakar$^{1}$ and Re'em Sari$^{2}$ }
\affil{1. Raymond and Beverly Sackler School of Physics \&
Astronomy, Tel Aviv University, Tel Aviv 69978, Israel\\
2. Racah Institute for Physics, The Hebrew University, Jerusalem
91904, Israel\\}

\begin{abstract}
The light from shock breakout of stellar explosions carries a wealth
of information on the progenitor and the explosion. The observed
breakout signature strongly depends on the velocity of the shock at
the time of breakout. The emission from Newtonian breakouts, typical
in regular core-collapse supernovae (SNe), was explored extensively.
However, a large variety of explosions result in mildly or ultra
relativistic breakouts, where the observed signature is unknown.
Here we calculate the luminosity and spectrum produced by
relativistic breakouts. In order to do so we improve analytic
description of relativistic radiation mediated shocks and follow the
system from the breakout itself, through the planar phase and into
the spherical phase. We limit our calculation to cases where the
post breakout acceleration of the gas ends during the planar phase
(i.e., the final gas Lorentz factor $\lesssim 30$). We find that
spherical relativistic breakouts produce a flash of gamma-rays with
energy, $E_{bo}$, temperature, $T_{bo}$, and duration,
$t^{obs}_{bo}$, that provide the breakout radius ($\approx 5R_\odot
(t_{bo}^{obs}/ 10 ~{\rm s}) \left(T_{bo}/50 ~{\rm keV}\right)^{2}$)
and Lorentz factor ($\approx T_{bo}/50 ~{\rm keV}$) and that always
satisfy $ (t_{bo}^{obs}/ 20 ~{\rm s}) \sim \left(E_{bo}/10^{46} {\rm
~erg~}\right)^{1/2} \left(T_{bo}/50 ~{\rm keV}\right)^{-2.68}~$. The
breakout flare is typically followed, on longer time scales, by
X-rays that carry a comparable energy. We apply our model to a
variety of explosions, including Ia and .Ia SNe, AIC, energetic SNe
and gamma-ray bursts (GRBs). We find that all these events produce
detectable gamma-ray signals, some of which may have already been
seen. Some particular examples are: (i) Relativistic shock breakouts
provide a natural explanation to the energy, temperature and time
scales, as well as many other observed features, of {\it all} low
luminosity GRBs. (ii) Nearby broad-line Ib/c (like SN 2002ap) may
produce a detectable $\g$-ray signal. (iii) Galactic Ia SNe produce
detectable $\g$-ray flares, if their progenitors are single
degenerate. We conclude that relativistic shock breakout is a
generic process for the production of gamma-ray flares, which opens
a new window for the study and detection of a variety of stellar
explosions.
\end{abstract}

\section{Introduction}
The first observable light from any stellar explosion is generated
by the breakout of the shock that traverse the star. In most stellar
objects the density drops sharply near the stellar edge. The shock
that encounters this density drop accelerates, leaving most of its
energy behind. Therefore, it breaks out at a velocity that is
significantly higher than the typical shock velocity in the stellar
interior, carrying only a small fraction of the total explosion
energy. In typical core-collapse supernovae (SNe), where the total
explosion energy is $\sim 10^{51}$ erg and the progenitor is not a
compact object, the velocity of the shock at breakout is Newtonian.
For example in a red supergiant progenitor the breakout velocity is
in the range $5,000-10,000$ km/s, while in a Wolf-Rayet progenitor
it is $\sim 30,000-100,000$ km/s. In cases where the explosion is
more energetic or the progenitor is a compact object, the breakout
can become mildly or even ultra relativistic.

Before its breakout, the shock is dominated by radiation. During
Newtonian breakout all the photons that are contained in the shock
transition layer are escaping the system and are seen as a short and
bright flash. The properties of Newtonian shock breakout flashes
were explored by many authors
\citep[e.g.,][]{Colgate74,Falk78,Klein78,Imshennik81,Ensman92,MatznerMcKee99,Katz10,Nakar10,Katz11}.
These properties depend mostly on breakout radius and on the shock
velocity at the layer that dominates the breakout emission, $v_0$.
Among these properties the observed typical photon frequency,
denoted here as the radiation temperature $T_0$, can be especially
sensitive to the shock velocity. In relatively slow shocks, $v_0 <
15,000$ km/s, the radiation in the shock breakout layer is in
thermal equilibrium and $T_0 \propto v_0^{1/2}$, producing typically
a spectrum that peaks in the UV. However, as was shown by
\cite{Weaver76}, in faster radiation mediated shocks the radiation
is away from thermal equilibrium in the immediate shock downstream,
and as a result the temperature in the shock transition layer is
much higher than the one obtained assuming thermal equilibrium. As a
result the breakout in fast Newtonian shocks, $v_0 \sim
30,000-100,000$ km/s, peaks in X-rays \citep{Katz10,Nakar10}.

The emission following the breakout can be divided into two
dynamically different regimes. At first, before the expanding gas
doubles its radius, the dynamics is planar while at later time the
evolution is spherical \citep{Piro10,Nakar10,Katz11}. The evolution
of the temperature and luminosity during the planar phase was
studied by \cite{Nakar10}, were we have shown that if the breakout
emission is out of thermal equilibrium at breakout, it remains out
of equilibrium also during the planar phase and thermal equilibrium
is achieved only during the spherical phase. The evolution during
the spherical homologous phase, was studied by many authors
\citep[e.g.,][]{Grassberg71,Chevalier76,Chevalier92,Chevalier08,Piro10,Rabinak10,Nakar10},
where all use the assumption (that in \citealt{Nakar10} we prove to
be adequate) of thermal equilibrium. All these studies considered
only explosions where the shock breakout velocities are Newtonian
and pairs are not significantly produced in the transition layer,
i.e., $\beta_0=v_0/c<0.5$ where $c$ is the speed of light.

A large range of cosmic explosions are expected to produce breakout
velocities that are either mildly relativistic or ultra
relativistic. A few examples are a range of white dwarf explosions
such as Ia SN \citep{Piro10}, .Ia SN \citep{Shen10} and possibly
accretion induced collapse \citep{Dessart06,Piro2010AIC}. The energy
involved in most of these explosions is $\lesssim 10^{51}$ erg, but
the compact progenitor and relatively low ejecta mass results in
relativistic breakouts. Another example is extremely energetic
explosions of massive stars, with explosion energies in the range
$10^{52}-10^{53}$ erg. These were recently observed in growing
numbers, e.g., SN2005ap and SN2007bi
\citep[e.g.,]{Gal-Yam09,Quimby09}, and are believed to be the
outcomes of various progenitor systems. Yet another example is
massive star explosions that are associated with long gamma-ray
bursts, where the breakout is ultra-relativistic. Relativistic
breakouts may also take place in choked GRBs, where the relativistic
jets fail to penetrate through the stellar envelope, and no regular
long GRB is produced. As we show here, this mechanism is a promising
source of all the observed low-luminosity GRBs.

The physical processes involved in producing the observed signature
of a relativistic shock breakout is different from the Newtonian
case in two fundamental ways. First, once the velocity of a
radiation mediated shock becomes large enough, $\beta_s > 0.5$, and
the temperature behind the shock exceeds $50$ keV, pair production
becomes the dominant source of leptons, which in turn are the main
source of photons via free-free emission. The exponential
sensitivity of the number of pairs to the temperature regulates the
rest frame temperature in the shock immediate downstream to an
almost constant value of $100-200$ keV, independent of the shock
Lorentz factor \citep{Katz10,Budnik10}. The gas becomes loaded with
pairs that dominate its optical depth, resulting in a temperature
dependent opacity per unit of mass. This is in sharp contrast to
Newtonian shocks, especially when $0.05<\beta_s<0.5$, where the
temperature of the shocked  fluid depends strongly on the shock
velocity but the opacity per unit of mass is constant. The second
major difference of relativistic explosions is the hydrodynamics
both before and after the breakout. A Newtonian shock that
propagates in a power-law decreasing density assumes the self
similar solution of \cite{Sakurai60} before breakout, and the gas
velocity is roughly constant after the breakout (it doubles by about
a factor of two). Once the shock becomes relativistic it follows the
solution of \cite{Johnson71} (see also \citealt{Tan01} and
\citealt{Pan06}) before the breakout and each fluid element
continues to accelerate significantly following the breakout. These
two differences, together with the usual relativistic effects such
as relativistic beaming, govern the observed emission, which as we
show here is very different than in the Newtonian case.

Yet another difference between Newtonian and relativistic radiation
mediated shocks is the physical width of the shock. In radiation
mediated shocks $\tau_s \sim c/v_s$, where $\tau_s$ is the total
optical depth (including pairs and Klein-Nishina effects) seen by a
photon going from the shock downstream towards the shock upstream.
In Newtonian radiation mediated shocks, where no pairs are produced,
the optical depth of a layer remains similar before it is shocked
and once it becomes the shock transition layer. Therefore, the shock
breaks out at the layer in which $\tau=\tau_s$, where $\tau$ is the
Thompson optical depth to the stellar edge {\it before} the shock
crossing (i.e., without pairs). However, in relativistic shocks,
where pairs are produced and Klein-Nishina effects may become
important, the  optical depth of a layer is changed significantly
when it is swept up by the shock and the relation between $\tau_s$
and $\tau$ is not trivial anymore. In order to find this relation
one needs to know the structure of the shock transition layer. The
structure of relativistic radiation mediated shocks in different
regimes was solved numerically by \cite{Budnik10} and
\cite{Levinson08}. The solution of \cite{Budnik10}, where a
significant number of photons is generated within the shock, is the
relevant one to the type of shocks we consider here
\citep{BrombergLevinson11}. \cite{Budnik10} also provide an
approximate analytic description of the shock structure as function
of the optical depth. We derive a more accurate analytic description
of the structure of the shock transition layer and use it to find
the value of $\tau$ at the point that the shock breaks out. We find
that while production of pairs results in a shock that breaks out of
the star at $\tau \ll 1$, the observed emission is always dominated
by the layer where $\tau \approx 1$. We therefore use the subscript
`$_0$' and the terminology `\bo shell' to denote the $\tau =1$
layer, which is not the actual layer where the breakout takes place,
but is the layer that dominates the observed breakout emission.

In this paper we calculate the evolution of the observed luminosity
and temperature from explosions in which the shock becomes mildly or
ultra relativistic, i.e., $\g_0 \beta_0 >0.5$. We follow the light
curve as long as the observed radiation is generated by gas that is
moving at relativistic velocities. At later times the radiation is
determined by Newtonian gas, which light curve was discussed in
\cite{Nakar10}. Our solution is limited to cases where the breakout
shell ends its post shock acceleration before it doubles its radius.
This limit is translated to a final (post-acceleration) Lorentz
factor of the \bo shell $\lesssim 30$. We also assume that the
opacity of the progenitor wind is negligible and do not affect the
breakout emission. Our calculations are applicable to a wide range
of energetic and/or compact explosions including Ia and .Ia SNe,
energetic core collapse and pair instability SNe and possibly part
of the phase space of chocked GRB jets.

We find that typically, relativistic breakouts produce flashes of
gamma-rays that without further information (e.g., the burst
redshift) can be mistaken as cosmological gamma-ray bursts. These
results provide a closure to \cite{Colgate68}, who was the first to
predict that shock breakouts produce flares of gamma-rays, even
before the detection of gamma-ray bursters  was announced
\citep{Klebesadel73}. However, \cite{Colgate68} assumed that the gas
behind the shock is in thermal equilibrium, and the observed
$\g$-rays are due to blue shift of ultra-relativistic gas, brought
to high Lorentz factors by shock that accelerates indefinitely as it
approaches the stellar surface. It ignores deviation from thermal
equilibrium and the fact that acceleration stops once the shock
width is comparable to the distance to the stellar surface. As it
turned out, the deviation from thermal equilibrium increases the
photon's rest frame temperature and compensates for the limited
Lorentz factor of the gas, resulting in a burst of $\g$-rays. Since
Colgate's prediction, breakouts were suggested to produce bright
gamma-ray emission by several authors in the context of
low-luminosity GRBs \citep{Kulkarni98,
Tan01,Campana06,Waxman07,Wang07}. All these calculations assume
thermal equilibrium, underestimating the true radiation temperature.
\cite{Katz10} realized that the deviation from equilibrium,
discussed by \cite{Weaver76}, will lead to $\g$-ray breakouts. They
did not follow, however, the post shock dynamics and opacity
evolution to find the properties of the observed emission. Here we
provide, for the first time, a quantitative description of the
predicted light curve from relativistic breakouts. We show that
these light curves explain very well many of the observed features
of low-luminosity GRBs, and also, that a variety of other stellar
explosions are most likely producing a bright gamma-ray breakout
emission.

In \S \ref{sec dynamics} we discuss the pre and post breakout
dynamics. The dynamics is dominated by mass, energy and momentum
conservation and is therefore largely independent of the photon
number and spectrum, as long as the gas is optically thick. In \S
\ref{sec Temp and opacity} we calculate the temperature and opacity
evolution of the expanding gas. \S \ref{sec light curve} follows the
observed light curve from the breakout to the spherical phase. Our
results are applied in \S \ref{sec observed cases} to find the
observed signature of known and hypothesized explosions. An analytic
description of the structure of the transition layer in relativistic
radiation mediated shocks is presented in Appendix A.


\section{Dynamics}\label{sec dynamics}
\subsection{Pre-breakout dynamics}
Consider a relativistic blast wave that propagates near a stellar
edge pre-explosion mass density gradient of the form $\rho \propto
z^n$, where $z=(R_*-r)$, $R_*$ is the stellar radius and $r$ is the
distance from the star center. In compact stars, where shocks are
more likely to become relativistic, the envelope is radiative and
typically $n \approx 3$, which is the value that we use throughout
the paper. The shock acceleration stops when the width of the shock
is comparable to $z$ and the shock breaks out. As we show in
appendix A, this takes place in the shell where  the pre-explosion
optical depth to the stellar edge is $\tau \sim 0.01 \g_s$, where
$\g_s$ is the shock Lorentz factor. Namely, breakout takes place at
$z \sim 0.01 \g_s/(\rho \kappa_T)$, where $\kappa_T \approx 0.2 {\rm
~cm^2/g}$ is the Thompson cross section per unit of
mass\footnote{Throughout the paper we use this value, which is
appropriate for hydrogen free ionized gas. The dependence on the
value of $\kappa_T$ is weak and all our results are applicable also
to hydrogen rich ionized gas. Note that through most of the phases
discussed here the temperatures in the system are larger than 10
keV, where the gas is fully ionized regardless of its composition,
implying that the results are insensitive to metallicity.}. This
sets the maximal Lorentz factor of the shock and we refer to that
shell as the {\it \ou shell}. Unlike the Newtonian case, the
observed emission from relativistic breakout is not dominated by the
\ou shell. Instead it is dominated by the shell where the
pre-explosion optical depth is\footnote{This is the criterion for
the dominating shell as long as $\g_0 \lesssim 100$. As we discuss
below this criterion is always satisfied in the regime at which our
analysis is applicable} $\tau \sim 1$ (i.e., $z \sim 1/(\rho
\kappa_T)$), which we therefore refer to as {\it the \bo shell} and
denote its properties by the subscript `$_0$'. In case of properties
that evolve with time, this subscript refers to their value in the
\bo shell right after it is crossed by the shock. Given the above
assumptions the system is completely defined by the following
parameters of the breakout shell:
\begin{itemize}
\item $\g_0$: the Lorentz factor of the shock when it crosses the breakout shell.
\item $m_0$: the mass of the breakout shell.
\item $d_0$: the post-shock width (i.e., after compression) of the breakout shell in the lab frame.
\item $R_*$: the radius of the star.
\item $n$: the power law index describing the pre-explosion density profile, here we use $ n =
3$.
\end{itemize}
Other characteristics of the breakout shell can be calculated from
the above. For example, the initial thermal energy of the breakout
shell is $E_0\approx \g_0^2  m_0 c^2$.

The pre-breakout and post-breakout dynamics of a planar relativistic
blast-wave in a decreasing density profile was investigated by
\cite{Johnson71}, \cite{Tan01} and \cite{Pan06}. If the density
follows a power-law then the evolution is self similar both before
and after the breakout. Before breakout a relativistic shock
accelerates as $\g_i \propto \rho^{-\mu}$ where
$\mu=\left(\sqrt{3}-\frac{3}{2}\right) \approx 0.23$ and the
subscript `$_i$' indicates initial values (note that $\rho$ is the
pre-shock density). This propagation profile sets the entire
pre-explosion hydrodynamic profile. Conceptually, it is useful to
treat the whole expanding envelope as a series of successive shells.
The properties of any shell deeper than the \ou shell, can be
characterized by its mass, $m$:
\begin{itemize}
\item $\g_i=\g_0 (m/m_0)^{-\mu n \over n+1} \propto m^{-0.17}$ is the initial Lorentz factor.
\item $d_i \approx z/\g_i^2 \propto m^\frac{1+2n\mu}{(n+1)} \approx
    m^{0.6}$ is the width of a shocked shell at shock breakout, in the lab frame.
\item $d_i'=d_i \g_i \propto m^\frac{1+n\mu}{(n+1)} \approx
    m^{0.42}$ is the width of a shocked shell at shock breakout, in the shell rest frame.
\item $E_i \approx \g_i^2 mc^2 \propto   m^{\frac{1+n(1-2\mu)}{n+1}}\approx m^{0.65}$ is the internal energy of a shell at shock breakout, in the lab frame.
\item $t_i \approx \frac{d_i}{c} \g_i^2 \propto m^\frac{1}{n+1} \propto m^{0.25}$ is the lab frame time for the shell expansion, which is
also the time between the shock crossing and the breakout
\end{itemize}

\subsection{Post-breakout dynamics}\label{sec postBO dynamics}
Following breakout all the shells expand and accelerate. The
acceleration is faster than that of a freely expanding relativistic
shell as outer shells gain energy from inner ones. In this
subsection we follow the acceleration of shells with mass $m \gtrsim
m_0$. These shells remain opaque during the whole planar phase, also
after their pairs annihilate. The final Lorentz factor of less
massive shells depends on their evolving opacity and is discussed
shortly in the next section.

The acceleration during the planar phase, i.e., before a shell
doubles its radius, follows \citep{Johnson71,Pan06}:
\begin{equation}
    \g = \g_i \left(\frac{t}{t_i}\right)^\frac{\3-1}{2}\propto
    m^{-0.27}t^{0.37} ,
\end{equation}
where $t$ is the lab frame time (not to be confused with observer
time) measured since breakout. If the shell is optically thick at
the end of acceleration and acceleration ends during the planar
phase then the final Lorentz factor is
\begin{equation}\label{eq gf}
    \g_f=\g_i^{1+\3}\propto m^{-0.48}.
\end{equation}
This relation is general as it does not depend on the exact density
profile, as long as there is a large energy reservoir behind the
accelerating shell. The acceleration ends at
\begin{equation}\label{eq tf}
    t_f = t_i \g_i^{3+\3}\propto m^{-0.57} ~.
\end{equation}
Thus, more massive shells end their acceleration at earlier times
and lower Lorentz factors.

In this paper we restrict our treatment to cases where the \bo shell
ends its acceleration during the planar phase, i.e., $t_{f,0}<t_s$
where
\begin{equation}\label{eq: ts}
    t_s=\frac{R_*}{c}
\end{equation}
is the lab frame time of transition between the planar and spherical
phases. Since pre-shocked optical depth of the \bo shell is of order
unity (i.e., $\kappa_T \rho_0 z_0 \approx 1$) :
\begin{equation}\label{eq t0_ts_ratio}
   \frac{t_0}{t_s}= \frac{z_0}{R_*} \approx \left(\frac{R_*^2}{\kappa_T M_*}\right)^{\frac{1}{n+1}}=3\cdot 10^{-3} \M^{-0.25}
   \R^{0.5}
\end{equation}
where $M_x$ and $R_x$ are the progenitor star mass and radius in
units of $x$ solar masses and radii respectively. Thus, there is a
critical Lorentz factor, $\g_{0,s}$, for which $t_{f,0}=t_s$:
\begin{equation}\label{eq g0 planar acceleration}
   \g_{0,s} = 3.5 ~\M^{0.05} \R^{-0.1} .
\end{equation}
The corresponding critical final Lorentz factor of the \bo shell is:
\begin{equation}
   \g_{f,s} = \g_{0,s}^{\3+1} = 30 ~\M^{0.14} \R^{-0.27}
\end{equation}
If $\g_0 \leq \g_{0,s}$  then acceleration ends in the planar phase
and the corresponding final Lorentz factor is given by equation
\ref{eq gf}. Otherwise it ends during the spherical phase and
equation \ref{eq gf} is not valid anymore, a case that we treat
elsewhere.

Acceleration continues beyond the \bo shell, up to the \ou shell,
which satisfies $z \sim 0.01 \g_i/(\rho \kappa_T)$. This shell
achieves the maximal initial Lorentz factor:
\begin{equation}\label{eq gmaxi}
    \g_{max,i} \approx 2 \g_0^{0.85}
\end{equation}
Note that our calculation is limited to $\g_0<\g_{0,s} \ll 100$,
where $\g_0<\g_{max,i}$ and the entire above description is
consistent.


\section{Temperature and opacity}\label{sec Temp and opacity}
The rest frame temperature behind a radiation mediated shock with
$\beta_s>0.5$  is roughly constant\footnote{This is true as long as
the blackbody temperature that corresponds to the post shock energy
density is lower than $T'_i$ } \citep{Katz10,Budnik10}:
\begin{equation}
    T'_i \sim 200 keV
\end{equation}
The reason is that at these velocities the radiation behind the
shock is out of thermal equilibrium and is set by the ability to
generate photons. The main photon source in the shock is free-free
emission \citep{Weaver76}. Once the temperature behind the shock
rises above $\approx 50$ keV, pairs are generated and their number
becomes dominant over the number of electrons that are advected from
the upstream. These pairs significantly increase the rate of photon
generation. The strong dependence of the number of pairs on the
temperature in that range serves as a "thermostat" that sets the
temperature in the immediate shock downstream to be nearly constant,
regardless of the shock Lorentz factor. Although the radiation is
out of thermal equilibrium the photons and the pair loaded gas in
the immediate downstream are in Compton-pair equilibrium and  there
is a single gas-radiation temperature \citep{Budnik10}. This
property of relativistic radiation mediated shocks is dominating the
emission at the shock breakout, during the planar phase and the
relativistic part of the spherical phase.

The gas behind the shock is pair loaded and therefore its optical
depth is very high, preventing any significant leakage of photons,
even if the pre-shocked (unloaded) gas optical depth is below unity
(the pairs set a diffusion time that is much larger than the
dynamical time). On the other hand, the rest frame temperature in
the expanding gas drops, and with it the number of pairs and the
photon generation rate. Therefore, the number of photons in the
opaque expanding shells is roughly constant. The mildly relativistic
temperature and the large optical depth ensure that photons and
pairs interact many times (i.e., sharing energy, annihilation and
creation) over the dynamical time, thereby keeping the pairs-photon
gas in Compton pair equilibrium.

The radiation of the \bo shell is confined to the gas during its
expansion until the rest frame temperature drops enough so the pair
loading becomes negligible. During the expansion of a shell its rest
frame temperature falls as $T' \propto V'^{-\beta}$, where $V'$ is
the shell volume (in its rest frame) and $1+\beta$ is an effective
adiabatic index. The value of $\beta$ depends on temperature and it
can drop slightly (by up to 30\%) below $1/3$ because of pairs
production and annihilation. Here we neglect this small deviation
from the typical relativistic equation of state and approximate
$\beta=1/3$. During the planar phase the volume of a fluid element
grows as $V' \propto t/\g$. Therefore during acceleration $V'
\propto t^{(3-\3)/2}$ while after the acceleration ends $V' \propto
t$. During the spherical phase $V' \propto t^3$ implying:
\begin{equation}\label{eq Ttag}
    T' = T'_i \left\{ \begin{array}{lc}
                              \left(\frac{t}{t_i}\right)^{-\frac{3-\3}{6}} &~~~~ t<t_f \\
                              \g_i^{-1} \left(\frac{t}{t_f}\right)^{-1/3} &~~~~~~t_f<t<t_s \\
                              \g_i^{-1} \left(\frac{t_s}{t_f}\right)^{-1/3}\left(\frac{t}{t_s}\right)^{-1}& ~~~t_s<t
                            \end{array}\right.
\end{equation}

The optical depth of the \bo shell drops to unity once the pairs
density, $n_\pm$, is lower than the proton density, $n_p$ (and their
accompanied electrons), i.e., $n_\pm/n_p<1$. The initial value of
this ratio is  \cite[e.g.,][]{Svensson84,Budnik10}:
\begin{equation}
   \frac{n_{\pm,0}}{n_{p,0}}
   \approx\frac{\g_0 m_p c^2}{3kT_i'} \approx 10^3 \g_0
\end{equation}
Since pairs are in annihilation-creation equilibrium their density
drops exponentially with $T'$ at this temperature range and the
shell opacity is dominated by the electrons that were advected from
the upstream once its temperature is:
\begin{equation}
   T'_{th} \approx 50~keV ,
\end{equation}
where the dependence on $\g_0$ is very weak. Around this
temperature, the breakout shell, and any less massive shell, becomes
transparent and its radiation escapes. We define $t_{th,0}$ as the
time that the \bo shell becomes optically thin, i.e., its
$T'=T'_{th}$.

There are three critical times in the system, in which the dynamics
and radiation change their behavior. These are the acceleration and
transparency times of the \bo shell, $t_{f,0}$ and $t_{th,0}$
respectively, and the planar to spherical transition time, $t_s$.
The observed signal depends on the relative order of these three
time scales. Here we restrict our analysis to the case that
acceleration ends during the planar phase, i.e., $t_{f,0}<t_s$.
Under this assumption, equation \ref{eq Ttag} implies:
\begin{equation}\label{eq tth0}
   t_{th,0}  =  \left\{ \begin{array}{lcr}
                              t_0\left(\frac{T'_0}{T'_{th}}\right)^{3+\3} &~~~~\frac{T'_0}{T'_{th}}<\g_0 & (t_{th,0}<t_{f,0})\\
                              t_0\left(\frac{T'_0}{T'_{th}}\right)^{3}\g_0^{\3} &~~~~~~\g_0<\frac{T'_0}{T'_{th}},\left(\frac{t_s T_{th}^{'3}}{t_0 T_0^{'3}}\right)^\frac{\3}{3} & (t_{f,0}<t_{th,0}<t_s)\\
                              t_s \frac{T'_0}{T'_{th}} \g_0^\frac{\3}{3} \left(\frac{t_0}{t_s}\right)^\frac{1}{3}& ~~~~~~\left(\frac{t_s T_{th}^{'3}}{t_0 T_0^{'3}}\right)^\frac{\3}{3}<\g_0<\frac{T'_0}{T'_{th}}  & (t_s<t_{th,0})
                            \end{array}\right.
\end{equation}
Now, $T'_0/T'_{th} \sim 4$  and we consider only $\g_0<\g_{0,s}$,
which for any non-degenerate progenitor implies $\g_0 < \g_{0,s}
\lesssim 4$. Therefore, the \bo shell becomes transparent only after
acceleration ends, i.e., $t_{f,0}<t_{th,0}$. On the other hand, the
opacity drops quickly during the spherical phase, so equation
\ref{eq tth0} implies that the \bo shell becomes transparent during
the planar phase, or at the beginning of the spherical phase, namely
$t_{th,0} \lesssim t_s$. Therefore for a large range of realistic
explosion parameters $t_{f,0}<t_{th,0}<t_s$, which is the case that
we present in detail below. Finally, note that shells with
$\g_i>\g_0$ can be accelerated only as long as they are optically
thick, namely their $T'>T'_{th}$. Therefore, only shells with $\g_i
\lesssim \min\{4, \g_{0,s}\}$ satisfy the relation
$\g_f=\g_i^{1+\3}$, shells with higher $\g_i$ end their acceleration
before this relation is satisfied.


\section{Light curve}\label{sec light curve}
\subsection{Breakout emission}
The radiation in the expanding gas is trapped by pairs at the time
of breakout. In the previous section we have shown that for most
progenitors in the regime that we consider, the \bo shell becomes
transparent after it ended acceleration and before, or soon after,
the transition to the spherical phase, i.e., $t_{f,0}<t_{th,0}
\lesssim t_s$. Therefore we derive below the observed light curve in
that case. Note that the \bo shell is not the only shell that
becomes transparent during the planar phase. All the lighter shells
($m<m_0$) out to the the \ou shell, become transparent during the
planar phase as well . Shells that are deeper (and more massive and
slower) than the \bo shell remain opaque during the planar phase and
become transparent only during the spherical phase.

Since slower moving shells carry significantly more energy, $E_i
\propto \g_i^{-3.75}$, the breakout emission is dominated by the
energy confined to the \bo shell at $t_{th,0}$, when its local frame
temperature is $T'_{th}$:
\begin{equation}\label{eq Ebo}
   E_{bo}= E_0 \frac{\g_{f,0}}{\g_0} \frac{T'_{th}}{T'_0} \approx  \frac{1}{4} m_0 c^2 \g_0
   \g_f \approx 2 \cdot 10^{45} \R^2
   \g_f^\frac{1+\3}{2} ~,
\end{equation}
where we use  $\tau_0 \approx \kappa_T m_0/(4\pi R_*^2)=1$  to find
\begin{equation}\label{eq m0}
    m_0 \approx  4 \cdot 10^{-9} M_\odot \R^2 ~.
\end{equation}
The observed temperature of this emission is
\begin{equation}\label{eq Tbo}
   T_{bo}=T'_{th} \g_{f,0} \sim 50 \g_f ~keV ~,
\end{equation}
and it is smeared by light travel time effects over a duration that
is comparable to  $t_s^{obs}$:
\begin{equation}\label{eq tbo}
t_{bo}^{obs} \approx t_s^{obs} \approx \frac {R_*}{c \g_{f,0}^2}
\approx 10 \frac{ \R}{\g_{f,0}^2} ~{\rm s}~,
\end{equation}
where $t_s^{obs}$ is the time of transition from the planar to the
spherical phase as seen by the observer. Clearly, equations \ref{eq
m0}-\ref{eq tbo} are general and are applicable to any gas density
profile. In fact also equation \ref{eq Ebo} is applicable to any
sharply decreasing density gradient, even if it is not a power-law.
The reason is that $m_0$ does not depend on this  profile and  so
does the relation between $\g_0$ and $\g_{f,0}$ (equation \ref{eq
gf}), as long as acceleration ends during the planar phase and there
is a large energy reservoir behind the \bo shell. Together,
equations \ref{eq Ebo}-\ref{eq tbo} provide three generic
observables that depend on only two parameters $\g_{f,0}$ and $R_*$.
Thus, the energy, temperature and time scale of the breakout flare
are enough to determine $R_*$ and $\g_{f,0}$, and they also must
satisfy
\begin{equation}\label{eq tbo test}
   t_{bo}^{obs} \sim 20 ~{\rm s}  \left(\frac{E_{bo}}{10^{46} {\rm ~erg~}}\right)^{1/2} \left(\frac{T_{bo}}{50 ~{\rm keV}}\right)^{-\frac{9+\3}{4}}  ~,
\end{equation}
if the source of the flare is a quasi-spherical\footnote{$E_{bo}$,
$T_{bo}$, and $t_{bo}^{obs}$ depend only on emission from a region
within an angle of $\sim 1/\g_{f,0}$ with respect to the line of
sight} relativistic shock breakout. This relativistic breakout
relation can be used to test the origin of observed gamma-ray
flares.

During $t<t_{bo}^{obs}$ the observer receives also photons from
shells that are outer of the \bo shell, which have higher observed
temperature than $T_{bo}$. These photons generate a high energy
power-law at frequencies $T_{bo}<\nu$. Shells with
$\g_i<T'_i/T'_{th} \approx 4$ obtain their terminal Lorentz factor
before they become optically thin, implying that $\g_f =
\g_i^{\3+1}$ and $E \propto E_i \g_f/\g_i \propto \g_i^{-2.01}$
while $T = T'_{th} \g_f \propto \g_i^{\3+1}$. Therefore the
integrated spectrum of the breakout flare exhibit a high energy
power-law
\begin{equation}\label{eq Fnu upper}
    \nu F_\nu(\nu>T_{bo}) \propto \nu^{-0.74}
\end{equation}
up to $\nu=\min\{2,0.25 \g_{0}^{2.05}\}{\rm~MeV}$, which can be an
order of magnitude larger than $T_{bo}$. During
$t_{obs}<t_{bo}^{obs}(=t_s^{obs})$ the observer receives also
photons from the \bo shell itself after it becomes transparent.
During this phase adiabatic cooling dictates
\citep[e.g.,][]{Nakar10},  $T \propto t_{obs}^{-1/3}$ and $L \propto
t_{obs}^{-4/3}$, implying
\begin{equation}\label{eq Fnu lower}
    \nu F_\nu(\nu<T_{bo}) \propto \nu
\end{equation}
The range of the low energy power-law in rather limited
$T_{bo}>\nu>T_{bo}(c t_{th,0}/R_*)^{1/3}= 0.5 T_{bo} \g_0^{1/\3}
\M^{-1/12} \R^{1/6}$. Note that equation \ref{eq Fnu upper} and the
spectral range where the high and low energy power-laws are
observed, do depend on the specific density profile and are given
for $n=3$.

At $t_s^{obs}$ the observed luminosity is still dominated by light
emitted when the \bo shell become transparent, $\sim t_{th,0}$. This
emission fades quickly until the emission from the spherical phase
becomes dominant. The fading light curve can be easily derive in
case that $\g_{f,0} \gg 1$, since then the observed luminosity is
dominated by emission of the \bo shell at large angles
($>1/\g_{f,0}$). The luminosity decays then as $L \propto
t_{obs}^{-2}$ and the temperature as $T \propto t_{obs}^{-1}$
\citep{Kumar00,NakarPiran03}, until the emission from the spherical
phase becomes dominant.

\subsection{Spherical phase} \label{sec spherical}
During the spherical phase the radius of the expanding sphere is not
constant anymore. Instead, the radius of the shells is $r \propto v
t$ and the luminosity is dominated by photons that are leaking from
the shell that satisfies $\tau \sim c/v$. We refer to this shell as
the {\it \lu shell} and denote its properties with the superscript
`$~\widehat{ }~$' (see \citealt{Nakar10} for a detailed discussion),
such that, e.g., the \lu shell satisfies, by definition,
$\tauh=c/\vh$. The evolution during the spherical phase depends on
whether the \lu shell is relativistic or not and on the initial
temperature of the \lu shell. As long as  $\gfh \gg 1$ the initial
temperature of the \lu shell is $\sim 200$ keV and its dynamics is
well approximated by the relativistic self-similar solution. Shells
with $0.5 < \g_i \beta_i < 1$ also have an initial temperature of
$\sim 200$ keV but their dynamics is better described by Newtonian
approximation. Namely, shock propagation according to the self
similar solution of \cite{Sakurai60} and no significant acceleration
after shock breakout. The transition time between the relativistic
and Newtonian phases is calculated below and it takes place around:
\begin{equation}\label{eq tNW}
  t_{NW}^{obs} = \frac{R_*}{c} \g_{f,0}^{1.05}= t_s^{obs} \g_{f,0}^{3.05}
\end{equation}
During the Newtonian phase $t_{obs} \propto \vh^{-4}$. The end of
the phase in which the initial temperature of the \lu shell is
constant takes place when $\vh \approx 0.5$c,
 which is at $t_{obs} \approx 10 t_{NW}^{obs}$. Here we
present luminosity and temperature evolution during the spherical
phase up to that point ($\vh \approx 0.5$c). At later times the
temperature drops quickly until thermal equilibrium is obtained.
Therefore there is a sharp break in the temperature evolution at $10
t_{NW}$. The light curve evolution at $t>10 t_{NW}$ is covered by
\cite{Nakar10}.

\subsubsection{Relativistic phase ($t_s^{obs}<t_{obs}<t_{NW}^{obs}$)} The
relativistic velocity of the shells dictates $\tauh \sim 1$ and
since $\tau \propto m/r^2$ the mass of the \lu shell is $\mh \propto
t^{2}$. The photons from the \lu shell arrive to the observer at
$t_{obs} \approx t/\gfh^2$ and their arrival is distributed over a
comparable duration. Therefore the \lu shell satisfies $\mh \propto
t_{obs}^{0.69}$, $\gih \propto t_{obs}^{-0.12}$, $\gfh \propto
t_{obs}^{-0.33}$ and $\Eh_i' \propto \mh \gih \propto
t_{obs}^{0.57}$, while $\rh \propto t \propto t_{obs}\gfh^2 \propto
t_{obs}^{0.34}$ and equation \ref{eq Ttag} dictates $\Th' \propto
t_{obs}^{-0.36}$.
The observed luminosity and temperature are:
\begin{equation}
   \Lh =\frac{\Eh}{t_{obs}}=\frac{\gfh \Eh_i'  (\Th'/T_i')}{t_{obs}} =
   \frac{E_0 c}{R_*} \left(\frac{t_0}{t_s}\right)^{1/3} \g_0^{\frac{4\3+9}{3}}
    \left(\frac{t_{obs}}{t_s^{obs}}\right)^{-1.12}
\end{equation}

\begin{equation}
   \Th \approx 200 {\rm ~keV~} \left(\frac{t_0}{t_s}\right)^{1/3} \g_0^\frac{\3}{3}
    \left(\frac{t_{obs}}{t_s^{obs}}\right) ^{-0.68}
\end{equation}
The relativistic phase ends once $\gih \approx 1$ at a time given in
equation \ref{eq tNW}.

\subsubsection{Newtonian phase with constant initial temperature
($t_{NW}^{obs}<t_{obs}<10~t_{NW}^{obs}$)} The luminosity during this
phase is independent of the temperature since pairs are long gone.
It is therefore similar to the luminosity evolution of the spherical
phase following a Newtonian breakout, where:
\begin{equation}
   \Lh \propto t_{obs}^{-0.35}
\end{equation}
Therefore a significant flattening of the luminosity evolution is
expected at $t_{NW}$. From this point the luminosity decay is steady
until recombination and/or radioactivity start playing a role.

The initial volume of the \lu shell is proportional to its initial
width (all shells starts at $R_*$), $\dhat_i \propto t^{0.44}$. The
observed temperature evolves as:
\begin{equation}
   \Th \approx T'_i \left(\frac{\vh^3 t^3}{\dhat_i R_*^2}\right)^{-1/3}
   \propto t_{obs}^{-0.6}
\end{equation}
This slope is not very different than that of the relativistic
phase. Therefore there is no significant feature in the temperature
evolution at $t_{NW}$. Only at $t_{obs} \sim 10~t_{NW}$ the
temperature starts dropping rapidly until the \lu shell gains
thermal equilibrium.

\section{Predicted signal of various explosions}\label{sec observed cases}
We use the model developed in the previous section to calculate the
signal expected from known and hypothesized explosions. There is a
large number of explosions, such as various SNe types, in which the
bulk of the mass is moving in Newtonian velocities but the shock
accelerates a minute amount of mass to a relativistic breakout. Such
explosions are often observed mostly in the optical, where the
kinetic energy, $E$, that is carried by the bulk of the ejected
mass, $M_{ej}$, can be measured. Therefore we first provide the
characteristic properties of the breakout and following emission of
such explosions as a function of $E$, $M_{ej}$ and $R_*$. Then we
use these results to discuss the predicted signals from various
explosions.

Given $E$, $M_{ej}$ and $R_*$ the initial Lorentz factor of the \bo
shell is determined by following the shock acceleration from the
Newtonian phase ($\g_s \beta_s \propto \rho^{-0.19}$), to the
relativistic phase ($\g_s \beta_s \propto \rho^{-0.23}$):
\begin{equation}\label{eq g0b0}
    \g_0 \beta_0 \approx 2.6 ~\E^{0.62} \Mej^{-0.45} \R^{-0.35} .
\end{equation}
where $\E=E/(10^{52} {\rm~erg})$. The numerical coefficient is taken
to fit the numerical results of \cite{Tan01}, which simulate the
transition of the shock from Newtonian to relativistic. Equation
\ref{eq g0b0} provides a good approximation to the breakout Lorentz
factor for $\beta_0>0.5$. The initial energy of the \bo shell is
\begin{equation}
    E_0 \approx 5\cdot 10^{46} {\rm~ erg} ~\E^{1.25} \Mej^{-0.9} \R^{1.3}
    ,
\end{equation}
and its final Lorentz factor is
\begin{equation}\label{eq gf0}
    \g_{f,0} \approx 14 \E^{1.7} \Mej^{-1.2} \R^{-0.95}
\end{equation}
where we use the approximation\footnote{This approximation is good
for $\g_0 \beta_0 > 2$. For lower values of $\g_0 \beta_0$ it is
better to calculate $\g_{f,0}=\g_0^{1+\3}=(\g_0^2
\beta_0^2+1)^\frac{1+\3}{2}$ and plug it into equation \ref{eq
breakout properties}} $\g_0 \approx \g_0 \beta_0$. Figure \ref{fig
gf0} shows a color map of the value of $\g_{f,0}$ as a function of
ejected mass and stellar radius for different values of explosion
energies, where only the range where our results are valid,
$0.5<\gamma_0 \beta_0 < \g_{0,s}$, is colored. The figure also
includes the typical phase space location of various explosions in
that range. It is evident that breakouts of white dwarf explosions,
extremely energetic SNe and possibly low luminosity GRBs are all
within the relevant velocity range.

\begin{figure}[t]
  \includegraphics[width=15cm]{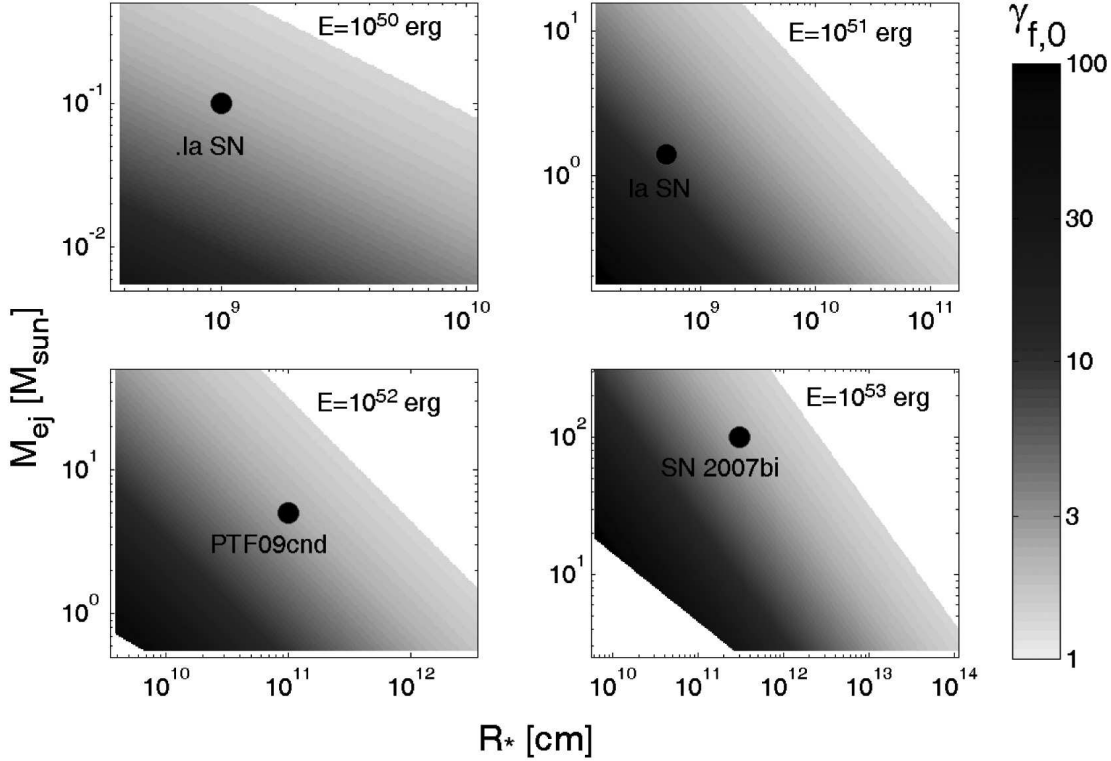}\\
  \caption{A color map of the value of $\g_{f,0}$ as function of ejected
mass and stellar radius for different values of explosion energies.
Only cases with $0.5<\gamma_0 \beta_0 < \g_{0,s}$ are colored. The
approximate phase space location of various explosions where the
breakout velocity falls in this range are marked as well.}\label{fig
gf0}
\end{figure}

Plugging equation \ref{eq gf0} into equations \ref{eq Ebo}-\ref{eq
tbo} we find the observed breakout emission:
\begin{align}\label{eq breakout properties}
    &E_{bo} \approx \frac{E_0 \g_{f,0}^\frac{3-\3}{2}}{4} \approx 6 \cdot 10^{46} {\rm~ erg} \E^{2.3} \Mej^{-1.65} \R^{0.7}\nonumber \\
    &\nonumber \\
    &T_{bo} \approx 50 {\rm~keV~} \g_{f,0} \approx 700 {\rm~ keV} \E^{1.7} \Mej^{-1.2} \R^{-0.95}\nonumber \\
     &\nonumber \\
    &t_{bo}^{obs} \approx t_{s}^{obs} = \frac{R_*}{c\g_{f,0}^2} \approx 0.06 {\rm ~s~} \E^{-3.4}  \Mej^{2.5}
    \R^{2.9}\\
     &\nonumber \\
    &L_{bo} \approx \frac{E_{bo}}{t_{bo}^{obs}} \approx 4 \cdot 10^{47} {\rm~ erg/s~}  \E^{5.1} \Mej^{-3.65} \R^{-1.85}\nonumber
\end{align}
Note that this equation is applicable only to a power-law density
profile (n=3) (unlike equations \ref{eq Ebo}-\ref{eq tbo} which are
general), due to the dependence of $\g_{f,0}$ on the density profile
in case that $E$, $M_{ej}$ and $R_*$ are the physical parameters.

The breakout emission dominates during the entire planar phase. At
the end of the planar phase ($t_{s}^{obs}$) the luminosity and
temperature drops quickly ($L \propto t_{obs}^{-2}$, $T \propto
t_{obs}^{-1}$ if $\g_{f,0} \gg 1$; see discussion in previous
section) to join the spherical emission, which is dominated by
relativistic ejecta until
\begin{equation}\label{eq tNW}
    t_{NW}^{obs} \approx 200 {\rm ~s~} \E^{1.8} \Mej^{-1.3},
\end{equation}
where we neglect a very weak dependence on $R_*$. The luminosity and
temperature at that time are
\begin{align}\label{eq L T tNW}
    &L(t_{NW}^{obs}) \approx 3 \cdot 10^{44} {\rm~ erg} \E^{0.9} \Mej^{-0.65} \R^{0.65}\nonumber \\
     & \\
    &T(t_{NW}^{obs}) \approx 0.2 {\rm~ keV} \E^{-3.15} \Mej^{2.3} \R^{1.9}\nonumber
\end{align}
where we ignore very weak dependence\footnote{In some explosions
$M_{ej} = M_*$ while in others $M_{ej} < M_*$. In any case the
observations in all the regimes we discuss here are almost
independent of $M_*$.} on $M_*$ and use the approximation $\g_0
\approx \g_0 \beta_0$. At $t_s^{obs}<t_{obs}<t_{NW}^{obs}$ the
luminosity shell is relativistic and $L \propto t_{obs}^{-1.12}$
while $T \propto t_{obs}^{-0.68}$. At
$t_{NW}^{obs}<t_{obs}<10t_{NW}^{obs}$ the luminosity shell is
Newtonian and $L \propto t_{obs}^{-0.35}$ while $T \propto
t_{obs}^{-0.6}$. The energy emitted per logarithmic scale in time,
$Lt_{obs}$, is rising during that last phase ($\propto
t_{obs}^{0.65}$), and most of the energy that is emitted in X-ray
(for typical parameters) is emitted around $t_{obs} \sim
10t_{NW}^{obs}$. This energy may be comparable to, or even larger
than, the breakout energy:
\begin{equation}\label{eq E10tNW}
    Lt_{obs}(10t_{NW}^{obs}) \approx 3 \cdot 10^{47} {\rm ~erg~} \E^{2.7} \Mej^{-1.9}
    \R^{0.65}
\end{equation}

At later time the luminosity evolution remains unchanged (until
recombination or radioactive decay become dominant), while the
temperature drops rapidly towards thermal equilibrium
\citep{Nakar10}.

\subsection{White dwarf explosions - Ia SN, .Ia SN and AIC}
There are several mechanisms that can result in a shock breakout
from a white dwarf. The most famous is type Ia SNe, where the whole
star explodes. Other theoretically predicted scenarios are .Ia SNe
\citep{Shen10}, where helium accreted in an AM CVn is detonated, and
accretion induced collapse (AIC) to a neutron star
\cite[e.g.,][]{Hillebrandt84,Fryer99,Dessart06} . All these
explosions are expected to produce a rather constant ejecta velocity
$(2E/M_{ej})^{1/2} \sim 10,000$ km/s, with explosion energy that
ranges between $10^{51}$ erg in type Ia SNe, where the whole star is
ejected ($M_{ej} \approx 1.4 M_\odot$), to $10^{49}-10^{50}$ erg in
the other explosions, where only $1-10$\% of the stellar mass is
ejected. There are also various observed SNe, which are probably
generated by a range of white dwarf explosions. Some possible
examples are SN 2005E, where $M_{ej} \approx 0.3 M_\odot$ and $E
\approx 4 \cdot 10^{50}$ erg \citep{Perets10}, and SN2010X where
$M_{ej} \approx 0.16 M_\odot$ and $E \approx 1.7 \cdot 10^{50}$ erg
\citep{Kasliwal10}.

Since $E/M_{ej}$ is similar for all these explosions, and the
dependence of $\g_0 \beta_0$ on $E$ (or $M_{ej}$) when this ratio is
constant, is very weak ($\propto E^{0.17}$), shock breakouts from
all these types of white dwarf explosions produce a similar
signature. The breakout Lorentz factor is $\g_0 \beta_0 \approx 1-3$
for a white dwarf radius $3-10 \times 10^8$ cm. The total energy
released in the breakout is $10^{40}-10^{42}$ erg and the typical
photon energy is $\sim$ MeV. The breakout pulse is spread over $\sim
1-30$ ms having a peak luminosity of $\sim 10^{44}$ erg/s. These
results for the breakout emission are different, especially in the
temperature prediction, than those of \cite{Piro10} that found a
type Ia breakout signal that peaks in the X-rays. The reason is that
they assume thermal equilibrium behind the shock, which is not a
good assumption for type Ia SNe \citep{Nakar10}, and that they
ignore relativistic effects.

Note that this signature is expected only if the line of sight to
the observer is transparent. In case that the explosion is triggered
by white dwarf mergers, the optical depth of the debris that
surrounds the exploding core can be high and the shock breakout will
take place at much larger radius than the white dwarf radius,
carrying out much more energy at a much lower velocity. For example,
\cite{Fryer10} find that a type Ia SN shock breakout in case of a
double degenerate merger take place at a radius of $\sim
10^{13}-10^{14}$ cm where the shock is Newtonian. Thus, a breakout
emission from type Ia SN can potentially differentiate between
single and double degenerate progenitor systems.

A flare of $10^{41}$ erg in MeV $\g$-rays can be detected easily if
it takes place in the Milky way and possibly also in the Magellanic
clouds. Such detection will provide a constrain on the on the exact
explosion timing as well as on the radius of the exploding white
dwarf and explosion energy. It may also shed light on the explosion
mechanism. This detection channel is extinction free and may be the
main detection channel in case that the burst take place in an
obscured Galactic environment. Given the estimated Milky way rate of
type Ia SNe ($\sim 0.01 {\rm ~yr^{-1}}$), the chance to detect such
a burst in the coming decade is low, but not negligible. The rate of
events such as SN 2010X is very hard to constrain observationally,
based on current optical surveys. In addition, there are no
observational constraints on the rate of events with energy lower
than $10^{50}$ erg, since their optical signature, where these
events are searched for, is too weak and evolve too fast for
detection. Similarly there are no tight theoretical constraints on
the rates of .Ia SNe and AICs. Since the gamma-ray shock breakout
signatures of all these events are similar and detectable for any
galactic event, it is worth looking for them in the current data of
satellites like BATSE, Konus-Wind, Swift and Fermi. In fact,
\cite{Cline05} find that the sub-sample of several dozen BATSE and
Konus-wind GRBs, those with duration shorter than $0.1$ s, shows a
significant anisotropy with concentration of events in the galactic
plane (but not towards the Galactic center).  This sample of events
may contain some already observed shock breakouts from white dwarfs.

\subsection{Broad line Ibc SNe}
Broad-line  Ibc SNe are thought to be explosions of Wolf-Rayet
progenitors, and in some cases are very energetic ($E \gg 10^{51}$
erg). Some Broad-line Ibc SNe are associated with GRBs (these are
discussed later), but also those that are not associated with GRBs
may produce a detectable $\g$-ray signal. For example SN 2002ap has
a total explosion energy of $\sim 4 \cdot 10^{51}$ erg and an
ejected mass of $\sim 2.5 M_\odot$ \citep{Maurer10}, implying a
mildly relativistic breakout with an energy output of $\sim 5 \times
10^{44} \R^{0.7}$ erg within $\sim 10 \R$ s. At a distance of about
7 Mpc the observed fluence on earth is $\sim 10^{-7} \R^{0.7} {\rm~
erg/cm^2}$. This fluence is detectable by sensitive gamma-ray
detectors even for $R_* = R_\odot$ (although no large field-of-view
detector with high sensitivity was operational in 2002). Note
however, that if a dense wind surrounds the progenitor then the
flare properties may be very different. The total emitted energy is
higher while the typical photon energy may fall either within or
below the observing band of soft $\g$-ray detectors.

\subsection{Extremely Energetic SNe}
Several types of extremely energetic SNe were detected in recent
years. SN 2007bi has shown a typical velocity of $12,000$ km/s and
$>50 M_\odot$ of ejecta implying $E \gtrsim 10^{53}$ erg
\citep{Gal-Yam09}. Most of the optical luminosity observed in this
SN is generated by radioactive decay. The radius of the progenitor
is unknown, but since there is no trace of either hydrogen or helium
its radius is most likely in the range, $R_* \sim 10^{11}-10^{12}$
cm. These values implies $\g_0 \beta_0 \approx 1$. Therefore the
breakout is expected to produce a $\sim 100$ keV flash that lasts
seconds to tens of seconds and carry an energy of $10^{44}-10^{46}$
erg.

\cite{Quimby09} reports on another type of extremely energetic SNe.
They find a typical velocity of $14,000$ km/s and $>5 M_\odot$ of
ejecta implying $E \gtrsim 10^{52}$ erg. The observed optical energy
in these explosions is not generated by radioactive decay and the
possible sources are interaction of the ejecta with matter that
moves more slowly or a central engine. In the former case, all the
energy is given to the ejecta during the explosion and a bright
shock breakout is expected. These explosions also do not show any
trace of hydrogen and Helium implying that their progenitor is also
rather compact. Assuming again $R_* \sim 10^{11}-10^{12}$ cm the
signature of the shock breakout from these events is similar to that
expected from SN 2007bi.

An energy output of $10^{44}$-$10^{46}$ in soft gamma-rays within a
minute is detectable out to a distance of 3-30 Mpc. Therefore the
detection of breakouts from such events is likely only if their rate
is larger than $10^{4} {\rm ~Gpc^{-3}~ yr^{-1}}$. The rate of
extremely energetic SNe is unknown, but it is most likely much lower
than that since otherwise they should have been detected in optical
searches in large bunbers. We therefore do not expect current
gamma-ray detectors to detect any of these SN types, although Nature
may surprise us.

\subsection{Low luminosity GRBs}
The engine of long GRBs is thought to be produced during the
collapse of a massive stellar core and thus to be launching
relativistic jets into a stellar envelope. \cite{Bromberg11a} show
that (at least in some) low-luminosity GRBs it is highly unlikely
that the jets successfully punch through the star and emerge in
order to produce the observed $\g$-ray emission. In that case what
can be the source of the observed gamma-rays?

There are several common features observed in all low luminosity
GRBs. All show $10^{48}-10^{50}$ erg of gamma-rays or hard X-rays
that are emitted in single pulsed light curves showing a spectra
that become softer with time (see \citealt{Kaneko07} and references
therein). They also show radio afterglows that indicate on similar
energy in mildly relativistic ejecta
\citep{Kulkarni98,Soderberg04,Soderberg06}. This energy is only a
small fraction of the total energy observed in the associated SNe
($10^{52}-10^{53}$ erg). The smooth light curves and the energy
coupled to a mildly relativistic ejecta motivated several authors to
suggest that the origin of this emission is shock breakouts
\citep{Kulkarni98,Tan01,Campana06,Waxman07,Wang07}. Later,
\cite{Katz10} have shown that the deviation from thermal equilibrium
can explain the observed $\g$-rays, compared to the expected X-rays
when equilibrium is wrongly assumed. However, the question whether
shock breakouts are indeed the source of low-luminosity GRBs
remained open, mostly since there was no quantitative model that
could answer whether shock breakouts can indeed explain even the
most basic observables, such as energy, temperature and time scales,
of these bursts. Below, we use our model to answer this question. In
section \ref{sec light curve} we have shown that this three observed
scales, $E_{bo}$, $T_{bo}$ and $t_{bo}^{obs}$, over-constrain the
physical parameters since they depend in the case of a
quasi-spherical relativistic breakout only on two parameters,
$\g_{f,0}$ and the breakout radius. Thus, in addition to finding the
value of these two physical parameters, the three observables must
satisfy the relativistic breakout relation (equation \ref{eq tbo
test}). This can be used as a test that any flare that originates
from quasi-spherical relativistic breakout must pass. This result is
generic as it is independent of the pre-breakout density profile (as
long as it is steeply decreasing). Below we apply this test to the
four observed low-luminosity GRBs

The four well studied low luminosity GRBs are divided into two pairs
with very different properties. GRBs 060218 and 100316D are soft and
very long. Their rather soft spectrum is best fitted at early time
with a cut-off around a peak energy of $\sim 40$ keV
\citep{Kaneko07,Starling11} and both emitted a total energy of $\sim
5 \times 10^{49}$ erg. Plugging these values into equation \ref{eq
tbo test} we find that if these bursts are flares from
quasi-spherical breakouts then they should have durations of $\sim
1500$ s, which is indeed comparable to the observed durations of
these two bursts. The breakout radius in this model is $\sim 5
\times 10^{13}$ cm and $\g_{f,0} \approx 1$ implying $\g_0 \beta_0
\approx 0.3-1$. In the case of GRB 060218, radio emission suggest
breakout velocities on the upper side of this range
\citep{Soderberg06}. The large radius inferred by the long duration
implies that the breakout is most likely not from the edge of the
progenitor star, but from an opaque material thrown by the
progenitor to large radii (possibly a wind) prior to the explosion
\citep{Campana06}. The fact that a spherical model provides a good
explanation implies that the suggested a-sphericity
\citep{Waxman07}, if exist, does not play a major role in
determining the burst duration. Note though, that it may play an
important role in shaping the breakout spectrum above and below
$T_{bo}$ and the entire light curve following the breakout flare.

The other two low-luminosity GRBs, 980425 and 031203, are relatively
hard, $T \gtrsim 150$ keV, and not particularly  long, $\approx 30$
s. Their hard spectrum made it difficult to explain these events as
shock breakouts, before it was realized that the radiation will be
away from thermal equilibrium\footnote{For example, \cite{Tan01},
which consider shock breakout as the source of energy of GRB 980425,
add an interaction of the expanding gas with external mass shell, at
a radius where the gas is optically thin, in order to reprocess the
gas energy into internal energy and radiate it non-thermally.
\cite{Wang07} suggested that the gamma-rays are produced by bulk
comptonization of low energy thermal photons during the breakout. A
process that does not work at the relevant temperatures due to
Compton loses \citep{Katz10}}. Plugging $10^{48}$ erg and $150$ keV,
the total energy and typical frequency of GRB 980425, into equation
\ref{eq tbo test} results in a breakout duration of $\sim 10$ s,
comparable to the observed one. The breakout radius in that case is
$\sim 6 \times 10^{12}$ cm and the final Lorentz factor is\footnote{
Although this value of the Lorentz factor is below the lower-limit
derived by \cite{Lithwick01} for GRB 980425, there is no
contradiction between the two studies. The reason is that
\cite{Lithwick01} derived their limits assuming that there is a hard
spectral power-law that extends to very high energies, similarly to
the one observed in long GRBs. However such power-law does not exist
in the case or relativistic breakout and therefore their constraints
are not applicable to the model that we discuss here.} $\g_{f,0}
\approx 3$. The associated supernova of GRB 980425, SN 1998bw, was
very energetic exhibiting $M_{ej} \approx 15 M_\odot$ and a kinetic
energy of $\approx 0.5 \times 10^{53}$ erg \citep{Iwamoto98}.
Interestingly, Setting $R_* \sim 6 \times 10^{12}$ and $M_{ej}=15
M_\odot$ in equation \ref{eq g0b0} we find that an explosion energy
of $E \approx 3 \times 10^{53}$ produces a breakout with the
observed properties of GRB 980425. This energy is slightly larger
than the one seen in the SN 1998bw, but it is close enough to
suggest that the  energy source of the SN explosion is also the one
of the shock breakout. The mild discrepancy between the two energies
may be a result of the actual stellar density profile (compared to
the power-law with $n=3$) or of a deviation of the explosion from
sphericity. Using the energy and typical frequency of GRB 031203, $5
\times 10^{49}$ erg and $> 200$ keV, equation \ref{eq tbo test}
predicts a duration $\lesssim 35$ s. The observed duration, $\sim
30$ s is consistent with this value and indicates that the actual
temperature in not much higher than 200 keV. The breakout radius in
that case is $\sim 2 \times 10^{13}$ cm and the final Lorentz factor
is $\g_{f,0}\approx 5$. The SN associated with GRB 031203, SN
2003lw, is similar in ejected mass and slightly more energetic than
SN 1998bw \citep{Mazzali06}. Setting $R_* \sim 2 \times 10^{13}$ and
$M_{ej}=15 M_\odot$ in equation \ref{eq g0b0} we find that an
explosion energy of $E \approx 10^{54}$ produces a breakout with the
observed properties of GRB 031203. Again this value is larger by
about an order of magnitude compared to the one observe in the
associated SN.

The breakout radii that we find for GRBs 980425 and 031203 are
larger than those of typical Wolf-Rayet stars, the probable
progenitors of GRB-SNe. This implies that either the breakout is not
from the edge of the progenitor star (e.g., from a wind) or that
low-luminosity GRB progenitors are larger than commonly thought
(which may help chocking the relativistic jet). Finally, a
prediction of relativistic shock breakouts is a significant X-ray
emission, that can be comparable in energy to that of the breakout
$\g$-ray flare. The X-rays are emitted during the spherical phase,
over time scale that is significantly longer than that of the
breakout emission. GRB 031203 has shown a bright signal of dust
scattered X-rays, which indicates on $\sim 2$ keV emission with
energy comparable to that of the gamma-rays, during the first $1000$
s after the burst \citep{Vaughan04,FengFox10}. These values fit the
predictions of a breakout model. For example, if the breakout is
from a stellar surface at $R_* = 2 \times 10^{13}$ and $M_{ej}=15
M_\odot$, then equations \ref{eq tNW}-\ref{eq E10tNW} predict a
release of $\sim 10^{50}$ erg at $\sim 2$ keV within the first hour
after the burst.

To conclude, we find that the energy, temperature, and time scales
of all four low-luminosity GRBs are explained very well by shock
breakout emission and that all of them satisfy the relativistic
breakout relation between $E_{bo}$, $T_{bo}$ and $t_{bo}^{obs}$
(equation \ref{eq tbo test}). These observables are largely
independent of the pre-shocked density profile and are therefore
applicable also for a breakout from a shell of mass ejected by the
progenitor prior to the explosion and possibly to a breakout from a
wind. Breakout emission can also explain many other observed
properties, such as the smooth light curve, the afterglow radio
emission, the small fraction of the explosion energy carried by the
prompt high energy emission, and the delayed bright X-ray emission
seen in GRB 031203. We therefore find it to be very likely that
indeed all low-luminosity GRBs are relativistic shock breakouts.

\subsection{GRB 101225A}
GRB 101225A is a peculiar explosion showing an $\sim$hour long
smooth burst of gamma-rays and X-rays followed by a peculiar IR-UV
afterglow, which is consistent with a blackbody emission. Its origin
is unclear. \cite{Thone11} argue for a cosmological origin and
estimate its redshift to be $\approx 0.3$, while \cite{Levan11}
argue that its origin is local. Here we do not attempt to determine
which, if any, of the two views is correct. Instead, given the
similarity of the high energy emission of GRB 101225A to some
low-luminosity GRBs, we ask whether shock breakout can be the source
of this emission, assuming that the burst is cosmological and that
the redshift estimate of \cite{Thone11} is correct.

We test if its observed properties satisfy equation \ref{eq tbo
test}, which with a total energy of  $\sim 10^{51}$ erg and
gamma-ray spectrum that peaks at $\sim 40$ keV predicts a duration
of $\sim 10^4$ s. This value is consistent with the constraints on
the actual duration of the flare \citep{Thone11}. The resulting
breakout radius is $\sim 3 \times 10^{14}$ cm. Interestingly
\cite{Thone11} present a model that attempts to explain the IR-UV
afterglow  by an expanding opaque shell with velocities at least as
high as $\beta \approx 1/3$ and initial radius of $2 \times 10^{14}$
cm. A breakout of a shock that accelerates such a shell produces a
high energy signal that is similar to the observed one. We therefore
conclude that if \cite{Thone11} redshift estimate is correct, then a
relativistic shock breakout probably provides the simplest
explanation for the high energy emission.

\subsection{Long gamma-ray bursts}
In the colapsar model of long GRBs the relativistic jets pierce
through the envelope and emerge in order to produce the observed
gamma-ray emission. During their propagation the jets drive mildly
relativistic bow shocks into the envelope and inflate a cocoon that
collimates the jets to an even narrower angle than their initial
launching opening angle
\citep[e.g.,][]{MacFadyen99,Matzner03,Morsony07,Mizuta09}. Recently,
\cite{Bromberg11b} derived an analytic description of this
interaction while the jet is propagating deep in the envelope,
before it encounters the sharp density gradient near the stellar
edge.

The term shock breakouts from GRBs is used in the literature to
describe several different phenomena. Here, similarly to the rest of
the paper, we focus on the emission that is escaping from the
breakout layer after the forward shock, which is driven by the jet
head and the cocoon into the stellar envelope, breaks out. The
structure of the shock that is driven into the stellar envelope is
highly non spherical but at the time of the size of causally
connected regions is small an a local spherical approximation is
appropriate. Thus, the theory discussed above, together with the
model presented in \cite{Bromberg11b} and the results seen in
numerical simulations \citep[e.g.,][]{Mizuta09}, can be used to
obtain a rough idea of the expected breakout signal.

We consider a typical GRB jet with a half-opening angle
$\theta_0=0.1$ rad and a total luminosity of $10^{50}$ erg/s. The
jet is launched continuously into a $5 M_\odot$ progenitor with a
radius of $5 R_\odot$. At the point that the jet reaches the steep
density drop (i.e. $r \sim R_*/2$) its velocity approaches the speed
of light and it is narrowly collimated to an angle of $\sim
\theta_0^2$ \citep{Bromberg11b}. The opening angle of the cocoon at
this point is $\approx \theta_0$. From that point the jet head and
the cocoon shock starts accelerating towards the stellar edge and to
expand sideways. Numerical simulations show that the head is
expanding sideways, becoming comparable in size to the cocoon which
does not spread significantly \citep[e.g., figure 17 in
][]{Mizuta09}. Once the fractional distance of the shock to the
edge, $z/R_*$, is comparable to the shock opening angle, in our case
$\sim \theta_0=0.1$, then the shock lose causal connection with the
sides and it can be approximated by the spherical solution of an
accelerating blast wave \citep{Johnson71}. Taking
$\g_s\beta_s(z=0.1R_*) \approx 1$ and using the relation
$\g_s\beta_s \propto z^{-0.69}$ with equation \ref{eq t0_ts_ratio},
we obtain $\g_0 \approx 10$. This value is larger than $\g_{0,s}$ so
the \bo shell accelerates also after the transition to the spherical
phase. We do not provide here the solution to such case, but the
initial energy in the \bo shell is $\sim 10^{48} (R_*/5R_\odot )^2$
erg and its energy increases during acceleration by at most an order
of magnitude. Assuming that there is no significant wind surrounding
the progenitor, this energy is released by a short, $\sim$ ms, pulse
of $>$MeV photons.

The main distinguishable feature of the breakout pulse is its harder
spectrum and low fluence compared to the total burst emission. The
pulse luminosity depends strongly on the progenitor size. If the
progenitor is rather large, $R_* \gtrsim 5 R_\odot$, then the
breakout can be detected by Swift and Fermi up to a redshift $\sim
0.1$. If it is much smaller, then the breakout is too faint for
detection. Even in cases that the breakout is detectable, it may be
difficult to separate it from the GRB itself. Nevertheless, it is
worth looking for an initial hard spike in nearby GRBs.
Interestingly, there is a sub-sample of BATSE GRBs that show an
initial short hard spikes followed by a softer burst
\citep{Norris06}.


\section{Summary}\label{sec summary}
We calculate the emission from  mildly and ultra relativistic
spherical shock breakouts. First, we determine the pre and post
breakout hydrodynamical evolution. This requires knowing the shock
width, which we obtain by deriving an improved analytic description
to the structure of the transition layer in relativistic radiation
mediated shocks, based on the results of \cite{Budnik10}. After
obtaining the hydrodynamical evolution we follow the temperature and
opacity evolution to find the observed light curve. Our calculations
are applicable to stellar explosions with shock velocity larger than
0.5c, in which the breakout shell ends its acceleration during the
planar phase, i.e., its final Lorentz factor $\lesssim 30$ for
typical stars. We consider only cases where the progenitor wind has
no effect on the observed emission.

A relativistic radiation mediated shock brings the gas to a roughly
constant rest frame temperature, $\sim 200$ keV, and loads it with
pairs \citep{Katz10,Budnik10}. Photons remain confined to the
post-shock expanding gas as long as pairs keep it optically thick. A
significant number of photons are released towards the observer only
from optically thin regions, which has a gas rest frame temperature
$\lesssim 50$ keV (low pair load) and unloaded pair gas opacity,
$\tau$, smaller than unity. Our solution of the shock structure
shows that the shock width is significantly smaller than $\tau = 1$,
implying that the shock accelerates also shells with $\tau \leq 1$,
which become transparent once their pairs annihilate. Among these
shells the one that carries most of the energy is the most massive
one, i.e., $\tau \approx 1$, which we refer to as the \bo shell. In
the regime that we consider the \bo shell becomes transparent during
the planar phase (i.e., before its radius doubles) and after it
achieves its final Lorentz factor. Light travel time and
relativistic effects dictate that the emission of the \bo shell,
once it become transparent, dominates during the entire planar
phase. Slower shells release their photons only during the spherical
phase. The processes described above produce the main following observables:\\
(i) {\it A $\g$-ray flare with a typical photon energy that ranges
between $\sim 50$ keV and $\sim 2$ MeV. The flare typically contains
a small fraction of the explosion energy and it may be significantly
shorter than the progenitor light crossing time.} This flare is
generated by the \bo shell when it becomes transparent.\\
The Energy, temperature and duration of the flare depend only on two
parameters, $\g_{f,0}$ and the breakout radius. Thus, in addition to
providing the value of these two physical parameters, the three
observables must satisfy the relativistic breakout relation
(equation \ref{eq tbo test}). This relation can be used to  test if
an observed flare may have been produced by a quasi-spherical
relativistic breakout. Note, that our calculation of the breakout
flare energy, temperature and duration are independent of the exact
profile of the pre-breakout density, as
long as it is steeply decreasing. \\
(ii) Due to light travel time effects, the breakout flare contains
photons from faster and lighter shells, producing a high energy
power-law spectrum, $\nu F_\nu \propto \nu^{-0.74}$. It also
contains photons emitted by the \bo shell after it becomes
transparent and cools adiabatically, producing a low-energy
power-law spectrum, $\nu F_\nu \propto \nu$. Both power-laws are
limited to about one order of magnitude in frequency.\\
(iii) The flare ends with a sharp decay at a time that coincides
with the transition to the spherical phase. If the breakout is ultra
relativistic (i.e., $\g_{f,0} \gg 1$) then $L \propto t_{obs}^{-2}$
and $T \propto t_{obs}^{-1}$, until the spherical phase emission
dominates.\\
(iv) After the flare ends, spherical evolution dictates a steady
decay of the luminosity.  At first, while relativistic shells
dominate the emission, $L \propto t_{obs}^{-1.1}$. Later, at
$t_{NW}^{obs}$, once Newtonian shells become transparent $L \propto
t_{obs}^{-0.35}$.
Note that during the latter phase the total emitted energy increases with time.\\
(v) The post flare temperature decays at first at a steady rate of
about $t_{obs}^{-0.6}$, generated during the spherical phase. A
sharp drop in the temperature is observed  when shells which were
not loaded by pairs during the shock crossing ($v_s \lesssim 0.5$)
dominate the emission. The drop is typically from the X-ray range to
the UV, and it is observed at about $10~t_{NW}^{obs}$.\\
(vi) The energy emitted before the steep temperature drop is often
comparable to that emitted during the breakout flare. Thus, {\it a
signature of relativistic breakout in many scenarios is a bright and
short $\g$-ray flare and a delayed X-ray emission with comparable
energy.}

We apply our model to a range of observed and hypothesized
explosions finding the following predictions:
\begin{itemize}
\renewcommand{\labelitemi}{$\bullet$}
\item {\bf Type Ia SNe and other white dwarf explosions}: A range of
white dwarf explosions, including observed Ia SNe and SNe 2005E \&
2010X as well as hypothesized .Ia SNe and AIC, are all predicted to
produce a similar breakout emission. A $\sim 1-30$ ms pulse of
$10^{40}-10^{42}$ erg composed of $\sim$ MeV photons. Such pulse
from a Galactic explosion is detectable, implying a low, yet not
negligible, chance to detect such flare from type Ia SNe in the
coming decade. The rate of the other types of explosions are not
known. Moreover there is almost no correlation between the
properties of the $\g$-ray flare and the luminosity of the optical
emission. Implying that Galactic breakouts from white dwarf
explosions may have already been observed, but not recognized as
such. In this context the suggested sub-class of very short GRBs
\citep{Cline05} may contain such events.

\item {\bf Broad-line Ib/c SNe}: Their progenitor is compact enough
and the energy of some is high enough to form mildly relativistic
breakouts. For example SN2002ap-like events are predicted to produce
a detectable breakout $\g$-ray signal in case that there is no thick
wind surrounding the progenitor.

\item {\bf Extremely energetic SNe}: SNe such as 2007bi \citep{Gal-Yam09} and
explosions of the type reported by \cite{Quimby09}, are energetic
enough to generate mildly relativistic breakouts in very massive
stars. These breakouts are likely to release $10^{44}-10^{46}$ erg
in soft $\g$-rays over seconds to minutes. Such flashes are
detectable in the nearby Universe, however these are rare events,
and the chance to have one in a detectable distance is low.

\item {\bf Low luminosity GRBs}: The origin of low-luminosity GRBs
is unknown. \cite{Bromberg11a} show that low-luminosity GRBs are
very unlikely to be produced by relativistic jets that pierce
through their host star, which is a necessary ingredient of the
collapsar model for long GRBs. In that case what can be the source
of the high energy emission? The smooth light curve and the mildly
relativistic ejecta that generate the radio emission of GRB 980425
lead \cite{Kulkarni98} and \cite{Tan01} to suggest that relativistic
shock breakouts are related to low luminosity GRBs. The detection of
a thermal component in GRB 060218 \citep{Campana06} and of
additional low luminosity GRBs with similar properties to GRB
980425, supported this suggestion \citep{Waxman07,Wang07},
especially after it was realized that the shock emission strongly
deviates from thermal equilibrium \citep{Katz10}. However, there was
no quantitative model of the breakout emission that includes
deviation from thermal equilibrium and gas-radiation relativistic
dynamics, which could test this suggestion. Moreover, it was unclear
whether shock breakouts can generate the energies seen in these
bursts (especially GRBs 980425 and 031203 due to their shorter
duration).

We find that relativistic shock breakouts can explain very well the
energy, temperature and time scales of the prompt emission of all
observed low-luminosity GRBs. Moreover, all four known bursts
satisfy the relativistic breakout relation between these three
observables (equation \ref{eq tbo test}). We also show that
relativistic breakouts provide a natural explanation to the low
gamma-ray luminosity compared to the total explosion energy and to
the hard to soft spectral evolution (including the late energetic
X-ray emission that inferred from dust echoes of GRB 031203). We
therefore find shock breakouts to be the most promising sources of
all low-luminosity GRBs. The connection between low-luminosity and
long GRBs (both share a relation to a similar type of SNe), suggests
that the energy source for the shock breakout are choked
relativistic jets that are launched by a GRB engine but fail to
punch through the star.

In this context we examine the high energy emission of GRB 101225A,
whose origin (local or cosmological) is undetermined yet, but its
prompt emission resemble low-luminosity GRBs. We find that if the
redshift determination of \cite{Thone11}, $z \approx 0.3$, is
correct then this burst also successfully passes the test set by
equation \ref{eq tbo test}, suggesting that the source of its high
energy emission is a mildly relativistic shock breakout at a
relatively large radius of $\sim 3 \times 10^{14}$ cm.

\item {\bf Long GRBs}: According to the popular view of long GRBs,
relativistic jets drill their way through the progenitor envelope
driving a forward shock into it. We estimate that a typical value of
the initial breakout Lorentz factor of this shock is $\g_0 \sim 10$.
Our theory cannot determine its final Lorentz factor, since it ends
its acceleration after the evolution enters the spherical regime.
Nevertheless, this value of $\g_0$ indicates that shock breakout
from long GRBs release $\sim 10^{48}(R_*/5R_\odot )^2$ erg in a
short pulse of $>$MeV photon. This pulse is too dim to be observed
in the typical redshift of long GRBs, but it may be observable in
nearby $z \sim 0.1$ GRBs.
\end{itemize}

We conclude that relativistic shock breakouts are a generic process
for the production of gamma-ray flares, which opens a new window for
the study and detection of a variety of stellar explosions.

We thank Omer Bromberg, Amir Levinson and Tsvi Piran for helpful
discussions. E.N. was partially supported by the Israel Science
Foundation (grant No. 174/08) and by an IRG grant. R.S. was
partially supported by ERC and IRG grants, and a Packard Fellowship.

\renewcommand{\theequation}{A-\arabic{equation}}
\setcounter{equation}{0}  
\section*{APPENDIX - The transition layer structure of relativistic radiation mediated shocks}\label{appendix A}  

Here we derive an analytic description of the transition layer of
relativistic radiation mediated shocks, based on the results of
\cite{Budnik10}. We use mostly similar notations to those of
\cite{Budnik10} and carry out the calculation in the shock frame,
where the flow is assumed to be in a steady state. We approximate
the gas-radiation in the transition region as two counter flows. One
is a flow of photons with energy $\sim m_ec^2$ that are generated in
the immediate downstream and flow towards the upstream at a drift
velocity $\sim c$. The other is a gas flow that is moving towards
the downstream. This is a flow of the fluid that is shocked, which
is moving with a Lorentz factor $\Gu$ in the far upstream (this is
the shock Lorentz factor) and is slowed down in the transition
region to $\be=1/3$ (i.e., $\G \approx 1$). The downstream moving
fluid is composed of protons with a proper density $n$ and pairs and
photons with a proper density $n_{\pm,\gamma}$. The photons in this
flow were part of the photon flow that moves towards the upstream,
before they were scattered once by the fluid leptons and joined the
fluid motion towards the downstream\footnote{Protons and pairs are
assumed to be coupled by collective plasma processes on scales much
shorter than any scale on interest in the system and thereby can be
treated as a single fluid. The photons in this flow are having the
same downstream drift motion, but are not directly coupled to the
protons and leptons.}. We are interested in the shape of the
transition, i.e., $\G$ and the pair fraction as a function of
optical depth and physical length. Our final goal is to find the
shock width in the upstream frame in units of the pre-shock (i.e.,
with no pairs) upstream Thomson optical depth.

The shock is generated by the interaction of the two
counter-streaming flows. Every collision between upstream moving
photon and a downstream moving fluid lepton[photon] transfers energy
and momentum to the photon[created lepton] and makes it joining the
 downstream moving fluid. The temperature of the pairs and fluid
photons is $\sim \G m_e c^2$, as seen in the fluid rest frame. In a
steady state the flux of photons that are moving towards the
upstream, $n_{\gamma \rightarrow_{US}}$, is equal to the number flux
of pairs and photons that are moving with the fluid (assuming that
the later dominate over the proton number flux), i.e., $\ngus
\approx \G \be n_{\pm,\gamma}$. Now, since upstream moving photon
that is scattered is joining the downstream moving fluid, $d (\G
n_{\pm,\gamma}) = -\ngus d \tau_s$, where $\tau_s$ is the optical
depth for photons moving toward the upstream, defined to be
increasing in the upstream direction\footnote{Our definition of
$\tau_s$ is similar to $-\tau$ in the notation of \cite{Budnik10}}.
Defining $\x \equiv n_{\pm,\gamma}/n$ and assuming $\G^2
>> 1$ (i.e., $\be \approx 1$), we obtain:
\begin{equation}\label{}
    \frac{d\x}{d\tau_s}=-\x .
\end{equation}

\begin{figure}[h!]
\begin{center}
 \includegraphics[width=0.7\textwidth]{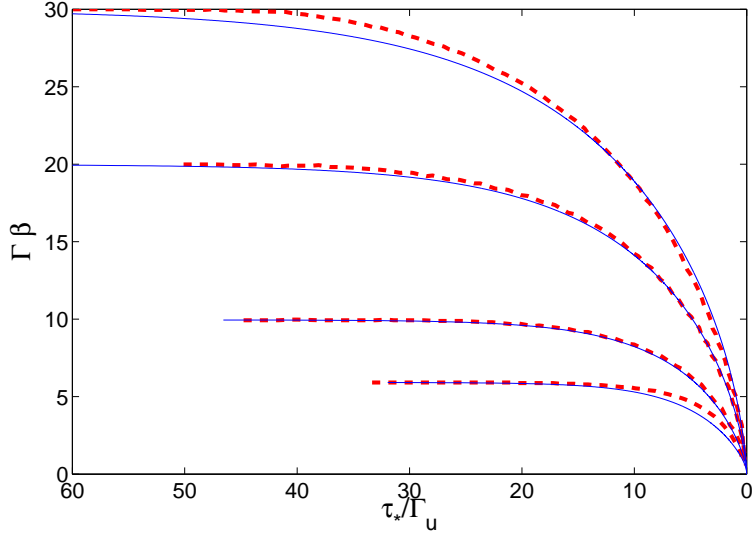}
\end{center}
\caption{Comparison of the shock structure for four values of $\G_u$
($6, 10, 20$ \& $30$), obtained by integrating over equation \ref{eq
dtaustar_dgamma} (using $\tau_* = \tau_s
\frac{\sigma_T}{\sigma_{KN}(1.4 \G^2)}$), [{\it thin solid line}],
to the numerical results of \cite{Budnik10} (their figure 6), [{\it
thick dashed line}]} \label{fig2}
\end{figure}

Number and momentum flux conservation imply:
\begin{equation}\label{}
n_u \Gu \be_u=n \G \be
\end{equation}
\begin{equation}\label{}
\Gu^2 \be_u n_u mp c^2=\G^2 \be n c^2 (m_p+4\x\G m_e)
\end{equation}
In the momentum equation we assume that the momentum flux of the
photon field that streams towards the upstream is negligible, which
is true as long as $\G^2 \gg 1$. We also assume $\x \gg 1$. These
equations imply:
\begin{equation}\label{}
\x =\frac{m_p}{4m_e} \frac{\Gu-\G}{\G^2}
\end{equation}
Therefore:
\begin{equation}\label{}
    \frac{d\G}{d\tau_s}=\frac{d\G}{d\x}\frac{d\x}{d\tau_s}=\G \frac{\Gu-\G}{2\Gu-\G}
\end{equation}
If we define $\tau_*$ as the Thompson optical depth of a photon
moving toward the upstream then $\tau_* \approx \tau_s
\frac{\sigma_T}{\sigma_{KN}(\G^2)}$, where $\sigma_T$ is the Thomson
cross-section and $\sigma_{KN}(\G^2)$ is the Klein-Nishina cross
section of electron at rest and photon with energy $\G^2 m_e c^2$.
Therefore
\begin{equation}\label{eq dtaustar_dgamma}
    d\tau_*=\frac{\sigma_T}{\G \sigma_{KN}(\G^2)} \frac{2\Gu-\G}{\Gu-\G} d\G
\end{equation}
Integrating this equation starting at the end of the transition
layer, i.e., $\G(\tau_s=0)=1$ results in the structure of the
transition layer. Comparison of this structure to the numerical
results of \cite{Budnik10} is presented in figure \ref{fig2}.

\begin{figure}[h]
\begin{center}
 \includegraphics[width=0.7\textwidth]{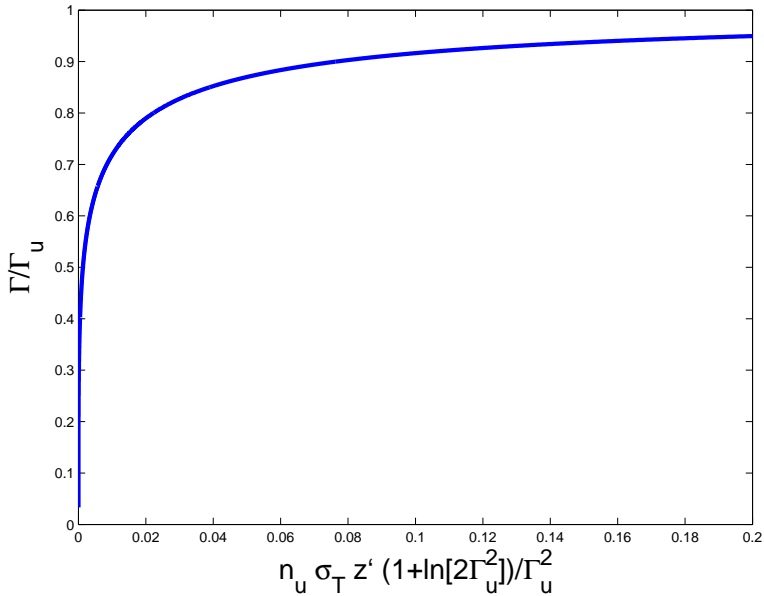}
\end{center}
\caption{The structure of the shock transition region of
relativistic radiation mediated shocks, presented in an optical
depth scale ($n_u \sigma_T z'(1+2\ln[\Gu^2])/\Gu^2$), which makes it
universal. i.e., the value of $\G/\Gu$ is independent of $\Gu$ (for
values larger than $1/\Gu$).} \label{fig3}
\end{figure}

In order to find the width of the shock in the upstream frame we use
the relation
\begin{equation}\label{}
    d\tau_*=\Gu n_u \x \sigma_T dz'
\end{equation}
where $z'$ is the length coordinate (positive towards the upstream)
in the shock frame. Therefore
\begin{equation}\label{}
    n_u \sigma_T dz'=\frac{4 m_e}{m_p}\frac{2\Gu-\G}{(\Gu-\G)^2\Gu}  \frac{\sigma_T}{\sigma_{KN}(\G^2)} \G d\G
\end{equation}
Integrating $d\G$ from the shock towards the upstream, the value of
$\G/\Gu$ (for values larger than $1/\Gu$) has a universal profile
(independent of $\G_u$) as a function of $n_u \sigma_T
z'(1+2\ln[\Gu^2])/\Gu^2$. This profile is presented in figure
\ref{fig3}. This profile implies that the shock width in unit of
pair unloaded Thomson optical depth in the shock frame is
roughly\footnote{Note that here we find the shock width in
pre-shock, {\it without pairs}, Thomson optical depth, which can be
directly related to physical width for a given upstream density.
\cite{Budnik10} find the shock width in units of  total Thomson
optical depth, {\it including pairs}, of a photon that crosses the
transition layer towards the upstream, $\tau_*$. As it turns out, in
both cases the width $\propto \G^2$ but the proportionality
coefficient is different (\citealt{Budnik10} find $\Delta \tau_*
\approx \G^2$).}, $\propto \G_u^2$. The value of the proportionality
coefficient is somewhat arbitrary and it depends on the value of
$\G/\Gu$ that defines the ``shock width". We chose $\G/\Gu=0.9$
obtaining $n_u \sigma_T z' \approx 0.01 \G_u^2$ (the value of the
logarithmic factor is $5-10$ for $\G_u =2-10$). Choosing a different
value of $\G/\Gu$, to define the shock width, changes the
coefficient, which in turn has a very weak effect on equation
\ref{eq gmaxi}. Now, since the length scale in the upstream frame,
z, is shorter by a factor $\G_u$ than in the shock frame (i.e.,
$z'=\Gu z$), and defining $\tau_u = n_u \sigma_T z$, we approximate
the shock width in units of Thomson optical depth of the pre-shock
upstream (i.e., with no pairs) as:
\begin{equation}\label{eq delta tau_u}
    \Delta \tau_u \sim 0.01 \G_u
\end{equation}
Note that our analytic description is applicable to shocks were
$\G_u^2 \gg 1$ and are therefore not directly applicable to mildly
relativistic shocks. We do expect however that pair creation in
these shocks, which increases the optical depth per proton in the
immediate downstream by a factor of the order of $m_p/m_e$, will
result in $\Delta \tau_u \ll1$ and that equation \ref{eq delta
tau_u} can be extrapolated to the mildly relativistic regime.


\begin{thebibliography}{}

\bibitem[{Bromberg}, {Mikolitzky} \& {Levinson}(2011)]{BrombergLevinson11}
{Bromberg}, O., {Mikolitzky}, Z., and {Levinson}, A. 2011, \apj,
733, 85,
  1101.4232.

\bibitem[{Bromberg}, {Nakar} \& {Piran}(2011)]{Bromberg11a}
{Bromberg}, O., {Nakar}, E., and {Piran}, T. 2011, In preperation.

\bibitem[{Bromberg} {\it et al.}\ (2011)]{Bromberg11b}
{Bromberg}, O., {Nakar}, E., {Piran}, T., and {Sari}, R. 2011,
Submitted to
  ApJ.

\bibitem[{Budnik} {\it et al.}\ (2010)]{Budnik10}
{Budnik}, R., {Katz}, B., {Sagiv}, A., and {Waxman}, E. 2010, \apj,
725, 63,
  1005.0141.

\bibitem[{Campana} {\it et al.}\ (2006)]{Campana06}
{Campana}, S. {\it et al.}\  2006, \nat, 442, 1008,
arXiv:astro-ph/0603279.

\bibitem[{Chevalier}(1976)]{Chevalier76}
{Chevalier}, R.~A. 1976, \apj, 207, 872.

\bibitem[{Chevalier}(1992)]{Chevalier92}
{Chevalier}, R.~A. 1992, \apj, 394, 599.

\bibitem[{Chevalier} \& {Fransson}(2008)]{Chevalier08}
{Chevalier}, R.~A. and {Fransson}, C. 2008, \apjl, 683, L135.

\bibitem[{Cline} {\it et al.}\ (2005)]{Cline05}
{Cline}, D.~B., {Czerny}, B., {Matthey}, C., {Janiuk}, A., and
{Otwinowski}, S.
  2005, \apjl, 633, L73, arXiv:astro-ph/0510309.

\bibitem[{Colgate}(1968)]{Colgate68}
{Colgate}, S.~A. 1968, Canadian Journal of Physics.~Vol.~46,
Supplement, p.476,
  46, 476.

\bibitem[{Colgate}(1974)]{Colgate74}
{Colgate}, S.~A. 1974, \apj, 187, 333.

\bibitem[{Dessart} {\it et al.}\ (2006)]{Dessart06}
{Dessart}, L., {Burrows}, A., {Ott}, C.~D., {Livne}, E., {Yoon}, S.,
and
  {Langer}, N. 2006, \apj, 644, 1063, arXiv:astro-ph/0601603.

\bibitem[{Ensman} \& {Burrows}(1992)]{Ensman92}
{Ensman}, L. and {Burrows}, A. 1992, \apj, 393, 742.

\bibitem[{Falk}(1978)]{Falk78}
{Falk}, S.~W. 1978, \apjl, 225, L133.

\bibitem[{Feng} \& {Fox}(2010)]{FengFox10}
{Feng}, L. and {Fox}, D.~B. 2010, \mnras, 404, 1018, 0908.0733.

\bibitem[{Fryer} {\it et al.}\ (1999)]{Fryer99}
{Fryer}, C., {Benz}, W., {Herant}, M., and {Colgate}, S.~A. 1999,
\apj, 516,
  892, arXiv:astro-ph/9812058.

\bibitem[{Fryer} {\it et al.}\ (2010)]{Fryer10}
{Fryer}, C.~L. {\it et al.}\  2010, \apj, 725, 296, 1007.0570.

\bibitem[{Gal-Yam} {\it et al.}\ (2009)]{Gal-Yam09}
{Gal-Yam}, A. {\it et al.}\  2009, \nat, 462, 624, 1001.1156.

\bibitem[{Grassberg}, {Imshennik} \& {Nadyozhin}(1971)]{Grassberg71}
{Grassberg}, E.~K., {Imshennik}, V.~S., and {Nadyozhin}, D.~K. 1971,
\apss, 10,
  28.

\bibitem[{Hillebrandt}, {Nomoto} \& {Wolff}(1984)]{Hillebrandt84}
{Hillebrandt}, W., {Nomoto}, K., and {Wolff}, R.~G. 1984, \aap, 133,
175.

\bibitem[{Imshennik}, {Nadezhin} \& {Utrobin}(1981)]{Imshennik81}
{Imshennik}, V.~S., {Nadezhin}, D.~K., and {Utrobin}, V.~P. 1981,
\apss, 78,
  105.

\bibitem[{Iwamoto} {\it et al.}\ (1998)]{Iwamoto98}
{Iwamoto}, K. {\it et al.}\  1998, \nat, 395, 672,
arXiv:astro-ph/9806382.

\bibitem[{Johnson} \& {McKee}(1971)]{Johnson71}
{Johnson}, M.~H. and {McKee}, C.~F. 1971, \prd, 3, 858.

\bibitem[{Kaneko} {\it et al.}\ (2007)]{Kaneko07}
{Kaneko}, Y. {\it et al.}\  2007, \apj, 654, 385,
arXiv:astro-ph/0607110.

\bibitem[{Kasliwal} {\it et al.}\ (2010)]{Kasliwal10}
{Kasliwal}, M.~M. {\it et al.}\  2010, \apjl, 723, L98, 1009.0960.

\bibitem[{Katz}, {Budnik} \& {Waxman}(2010)]{Katz10}
{Katz}, B., {Budnik}, R., and {Waxman}, E. 2010, \apj, 716, 781,
0902.4708.

\bibitem[{Katz}, {Sapir} \& {Waxman}(2011)]{Katz11}
{Katz}, B., {Sapir}, N., and {Waxman}, E. 2011, ArXiv e-prints,
1103.5276.

\bibitem[{Klebesadel}, {Strong} \& {Olson}(1973)]{Klebesadel73}
{Klebesadel}, R.~W., {Strong}, I.~B., and {Olson}, R.~A. 1973,
\apjl, 182,
  L85+.

\bibitem[{Klein} \& {Chevalier}(1978)]{Klein78}
{Klein}, R.~I. and {Chevalier}, R.~A. 1978, \apjl, 223, L109.

\bibitem[{Kulkarni} {\it et al.}\ (1998)]{Kulkarni98}
{Kulkarni}, S.~R. {\it et al.}\  1998, \nat, 395, 663.

\bibitem[{Kumar} \& {Panaitescu}(2000)]{Kumar00}
{Kumar}, P. and {Panaitescu}, A. 2000, \apjl, 541, L51,
arXiv:astro-ph/0006317.

\bibitem[{Levan} \& {Tanvir}(2011)]{Levan11}
{Levan}, A.~J. and {Tanvir}, N.~R. 2011, GRB Coordinates Network,
Circular
  Service, 11642, 1 (2011), 1642, 1.

\bibitem[{Levinson} \& {Bromberg}(2008)]{Levinson08}
{Levinson}, A. and {Bromberg}, O. 2008, Physical Review Letters,
100(13),
  131101, 0711.3281.

\bibitem[{Lithwick} \& {Sari}(2001)]{Lithwick01}
{Lithwick}, Y. and {Sari}, R. 2001, \apj, 555, 540,
arXiv:astro-ph/0011508.

\bibitem[{MacFadyen} \& {Woosley}(1999)]{MacFadyen99}
{MacFadyen}, A.~I. and {Woosley}, S.~E. 1999, \apj, 524, 262,
  arXiv:astro-ph/9810274.

\bibitem[{Matzner}(2003)]{Matzner03}
{Matzner}, C.~D. 2003, \mnras, 345, 575, arXiv:astro-ph/0203085.

\bibitem[{Matzner} \& {McKee}(1999)]{MatznerMcKee99}
{Matzner}, C.~D. and {McKee}, C.~F. 1999, \apj, 510, 379,
  arXiv:astro-ph/9807046.

\bibitem[{Maurer} \& {Mazzali}(2010)]{Maurer10}
{Maurer}, I. and {Mazzali}, P.~A. 2010, \mnras, 408, 947, 1006.1566.

\bibitem[{Mazzali} {\it et al.}\ (2006)]{Mazzali06}
{Mazzali}, P.~A. {\it et al.}\  2006, \apj, 645, 1323,
arXiv:astro-ph/0603516.

\bibitem[{Mizuta} \& {Aloy}(2009)]{Mizuta09}
{Mizuta}, A. and {Aloy}, M.~A. 2009, \apj, 699, 1261, 0812.4813.

\bibitem[{Morsony}, {Lazzati} \& {Begelman}(2007)]{Morsony07}
{Morsony}, B.~J., {Lazzati}, D., and {Begelman}, M.~C. 2007, \apj,
665, 569,
  arXiv:astro-ph/0609254.

\bibitem[{Nakar} \& {Piran}(2003)]{NakarPiran03}
{Nakar}, E. and {Piran}, T. 2003, \apj, 598, 400,
arXiv:astro-ph/0303156.

\bibitem[{Nakar} \& {Sari}(2010)]{Nakar10}
{Nakar}, E. and {Sari}, R. 2010, \apj, 725, 904, 1004.2496.

\bibitem[{Norris} \& {Bonnell}(2006)]{Norris06}
{Norris}, J.~P. and {Bonnell}, J.~T. 2006, \apj, 643, 266,
  arXiv:astro-ph/0601190.

\bibitem[{Pan} \& {Sari}(2006)]{Pan06}
{Pan}, M. and {Sari}, R. 2006, \apj, 643, 416,
arXiv:astro-ph/0505176.

\bibitem[{Perets} {\it et al.}\ (2010)]{Perets10}
{Perets}, H.~B. {\it et al.}\  2010, \nat, 465, 322, 0906.2003.

\bibitem[{Piro}, {Quataert} \& {Metzger}(2010)]{Piro2010AIC}
{Piro}, A., {Quataert}, E., and {Metzger}, B. 2010, in { AAS/High
Energy
  Astrophysics Division \#11}, volume~42 of { Bulletin of the American
  Astronomical Society}, 689.

\bibitem[{Piro}, {Chang} \& {Weinberg}(2010)]{Piro10}
{Piro}, A.~L., {Chang}, P., and {Weinberg}, N.~N. 2010, \apj, 708,
598,
  0909.2643.

\bibitem[{Quimby} {\it et al.}\ (2009)]{Quimby09}
{Quimby}, R.~M. {\it et al.}\  2009, ArXiv e-prints, 0910.0059.

\bibitem[{Rabinak} \& {Waxman}(2010)]{Rabinak10}
{Rabinak}, I. and {Waxman}, E. 2010, ArXiv e-prints, 1002.3414.

\bibitem[{Sakurai}(1960)]{Sakurai60}
{Sakurai}, A. 1960, Communications on Pure and Applied Mathematics,
13, 353.

\bibitem[{Shen} {\it et al.}\ (2010)]{Shen10}
{Shen}, K.~J., {Kasen}, D., {Weinberg}, N.~N., {Bildsten}, L., and
  {Scannapieco}, E. 2010, \apj, 715, 767, 1002.2258.

\bibitem[{Soderberg} {\it et al.}\ (2004)]{Soderberg04}
{Soderberg}, A.~M. {\it et al.}\  2004, \nat, 430, 648,
arXiv:astro-ph/0408096.

\bibitem[{Soderberg} {\it et al.}\ (2006)]{Soderberg06}
{Soderberg}, A.~M. {\it et al.}\  2006, \nat, 442, 1014,
  arXiv:astro-ph/0604389.

\bibitem[{Starling} {\it et al.}\ (2011)]{Starling11}
{Starling}, R.~L.~C. {\it et al.}\  2011, \mnras, 411, 2792,
1004.2919.

\bibitem[{Svensson}(1984)]{Svensson84}
{Svensson}, R. 1984, \mnras, 209, 175.

\bibitem[{Tan}, {Matzner} \& {McKee}(2001)]{Tan01}
{Tan}, J.~C., {Matzner}, C.~D., and {McKee}, C.~F. 2001, \apj, 551,
946,
  arXiv:astro-ph/0012003.

\bibitem[{Th{\"o}ne} {\it et al.}\ (2011)]{Thone11}
{Th{\"o}ne}, C.~C. {\it et al.}\  2011, ArXiv e-prints, 1105.3015.

\bibitem[{Vaughan} {\it et al.}\ (2004)]{Vaughan04}
{Vaughan}, S. {\it et al.}\  2004, \apjl, 603, L5,
arXiv:astro-ph/0312603.

\bibitem[{Wang} {\it et al.}\ (2007)]{Wang07}
{Wang}, X.-Y., {Li}, Z., {Waxman}, E., and {M{\'e}sz{\'a}ros}, P.
2007, \apj,
  664, 1026, arXiv:astro-ph/0608033.

\bibitem[{Waxman}, {M{\'e}sz{\'a}ros} \& {Campana}(2007)]{Waxman07}
{Waxman}, E., {M{\'e}sz{\'a}ros}, P., and {Campana}, S. 2007, \apj,
667, 351,
  arXiv:astro-ph/0702450.

\bibitem[{Weaver}(1976)]{Weaver76}
{Weaver}, T.~A. 1976, \apjs, 32, 233.

\end{thebibliography}

\end{document}